\documentclass[journal]{IEEEtran}
\usepackage{booktabs}
\usepackage{float}
\usepackage[dvips]{graphicx}
\usepackage{graphicx}
\usepackage{amssymb}
\usepackage{cite}
\usepackage{subcaption}
\usepackage{amsmath}
\usepackage{algorithm}
\usepackage{algorithmic}
\usepackage{multirow}
\makeatletter
\newcommand{\removelatexerror}{\let\@latex@error\@gobble}
\makeatother
\allowdisplaybreaks[4]
\usepackage{color}   
\makeatletter

\usepackage{makecell}
\makeatother
\usepackage[colorlinks = true,
            linkcolor = blue, 
            urlcolor = blue,
            citecolor = blue,
            anchorcolor = blue]{hyperref}

\begin{document}


\title{ Integrated Channel Estimation and Sensing for Near-Field ELAA Systems {via Low-Rank 
Tensor Decomposition}}

\author{Jionghui Wang, Jun Fang, Hongbin Li,~\IEEEmembership{Fellow,~IEEE} and Boyu Ning
\thanks{Jionghui Wang, Jun Fang and Boyu Ning are with the National Key Laboratory
of Wireless Communications, University of
Electronic Science and Technology of China, Chengdu 611731, China,
Email: jionghuiwang@std.uestc.edu.cn;JunFang@uestc.edu.cn;boydning@outlook.com}
\thanks{Hongbin Li is
with the Department of Electrical and Computer Engineering,
Stevens Institute of Technology, Hoboken, NJ 07030, USA, E-mail:
Hongbin.Li@stevens.edu}
}

\maketitle


\begin{abstract}
In this paper, we study the problem of uplink channel estimation for near-field orthogonal frequency division multiplexing (OFDM) systems, where a 
base station (BS), equipped with an extremely large-scale antenna array (ELAA), serves multiple 
users over the same time-frequency resource block. A non-orthogonal pilot transmission scheme is 
considered to accommodate a larger number of users that can be supported by ELAA systems without incurring an excessive amount of
training overhead. To facilitate efficient multi-user channel estimation, we express the received 
signal as a third-order low-rank tensor, which admits a canonical polyadic decomposition (CPD) model 
for line-of-sight (LoS) scenarios and a block term decomposition (BTD) model for non-line-of-sight (NLoS)
scenarios. An alternating least squares (ALS) algorithm and a non-linear least squares (NLS) 
algorithm are employed to perform CPD and BTD, respectively. Channel parameters 
are then efficiently extracted from the recovered factor matrices. By exploiting the
geometry of the propagation paths in the estimated channel,
users' positions can be precisely determined in LoS scenarios.
Moreover, our uniqueness analysis shows that the proposed tensor-based joint multi-user channel estimation 
framework is effective even when the number of pilot symbols is much smaller than
the number of users, revealing its potential in training overhead reduction. Simulation results demonstrate that the proposed method achieves 
markedly higher channel estimation accuracy than compressed sensing (CS)-based approaches.
\end{abstract}



\section{Introduction}
\subsection{Background}
Millimeter-wave (mmWave) and terahertz (THz) communications empowered by extremely large-scale antenna 
arrays (ELAA), which offer abundant spectrum resources and unprecedented spatial degrees of 
freedom (DoFs), are widely regarded as promising technologies for next-generation wireless communication systems
\cite{cui2023near,wang2024turorial,ning2023beamforming,3GPP_5G_Beamforming}. 
As the array aperture increases and the carrier wavelength decreases, the Rayleigh distance can extend to 
tens or even hundreds of meters, shifting the propagation characteristics from far-field planar wavefronts 
to near-field spherical wavefronts \cite{lu2021near,zhou2015spherical}. As a result, the channel response 
depends not only on the angle but also on the propagation distance. 
This additional range dimension enables distance-aware beam focusing and high spatial resolution, which are 
beneficial for both high-throughput communications and high-precision sensing \cite{cong2024near,zhang2025near}. 
Consequently, mmWave/THz systems with ELAA have emerged as a promising platform for integrated 
sensing and communication (ISAC), supporting various environment-aware applications, such as smart 
Internet of Things, autonomous driving, and extended reality \cite{wei2023integrated,liu2022integrated}.

In ISAC systems, channel estimation and sensing can be naturally integrated by reusing the same transmitted 
signals. Specifically, during the channel state information (CSI) acquisition phase, pilot signals can be 
exploited not only to estimate channel coefficients, but also to extract key physical parameters of the 
propagation paths, such as direction of arrival (DoA), propagation distance, and time delay. 
These parameters can subsequently be leveraged for channel reconstruction and user localization. 
However, the integrated channel estimation and sensing in near-field regime is challenging, 
since the spherical wavefront exhibits a nonlinear structure, where the angle and distance parameters are 
strongly coupled.

\subsection{Related Work}
Near-field channel estimation and source localization have attracted extensive attention in both wireless 
communication and radar sensing systems. 

For near-field channel estimation, the authors in \cite{cui2022channel} proposed a polar-domain 
sparse representation, where the dictionary is constructed by sampling over the 
two-dimensional (2D) angular-range domain. Based on the polar-domain codebook, near-field channel 
estimation can be formulated as a compressed sensing (CS) problem, which can be efficiently solved by a 
variety of methods \cite{cui2022channel,yue2024hybrid,yang2024near,prisharody2024near}. Nevertheless, 
joint sampling over this 2D domain results in a dictionary with a large number of atoms, leading to high 
computational complexity. In \cite{wang2024near}, authors showed that near-field channels exhibit 
a block-sparse representation on a specially designed unitary matrix. This new representation leads to a 
well-conditioned measurement matrix that is more amiable for CS of near-field channels. In addition 
to CS-based methods, tensor decomposition-based methods were developed for far-field mmWave/THz channel 
estimation \cite{zhou2016channel,zhou2017low} and later extended to near-field channel scenarios
\cite{cheng2025tensor} by exploiting inherent multi-dimensional channel structures.
 
For near-field source localization, classical subspace-based approaches, such as multiple signal 
classification (MUSIC) and estimation of signal parameters by rotational invariance techniques (ESPRIT), 
have been extended to a variety of near-field variants to achieve high-resolution angle and distance 
estimation. In \cite{huang1991near}, the authors proposed a 2D-MUSIC algorithm that jointly estimate 
the angles and ranges via a 2D-grid search. However, such a grid search is computationally intensive, 
especially when fine resolution is required. By exploiting the symmetric geometry of the antenna array, 
reduced-dimensional (RD) MUSIC and ESPRIT methods \cite{zhi2007near,ramezani2025massive} have 
been developed, where the angles are estimated via 1D-MUSIC/ESPRIT, and then the associated ranges can 
be calculated directly. 
Moreover, the near-field parameter estimation problem can also be formulated as a sparse reconstruction 	
task, which can be solved by CS-based methods \cite{teng2025near}.

More recently, several studies have investigated integrated sensing and channel estimation in 
near-field ISAC systems. In \cite{qiao2024sensing,lu2024near}, joint 
user localization and channel estimation are achieved through the two-stage OMP-based algorithm.  
A tensor-based method is designed in \cite{jiang2025near}, where the channel parameters are 
extracted from the tensor factor matrices using RD-MUSIC algorithm and correlation-based 
methods, enabling subsequent channel reconstruction and target localization. 

However, most of the aforementioned works rely on the assumption that orthogonal pilot sequences are 
employed across users. Under this assumption, the multi-user channel estimation problem can be decoupled 
into a set of independent single-user subproblems. In practical systems, especially when a large number of 
users are simultaneously served or when the pilot length is severely limited, pilot orthogonality cannot be 
guaranteed. In such scenarios, non-orthogonal pilots lead to inevitable inter-user interference, and the 
received signals corresponding to different users become intrinsically coupled. Consequently, the multi-user 
channel estimation problem can no longer be decomposed into independent subproblems, and existing single-user 
near-field estimation methods \cite{cui2022channel,yue2024hybrid,yang2024near,prisharody2024near,
wang2024near,cheng2025tensor,qiao2024sensing,lu2024near,jiang2025near} cannot be directly 
applied. Moreover, when
the pilot length is smaller than the number of users, the
received signal covariance matrix becomes rank-deficient, such
that the dimension of the signal subspace is smaller than the
number of users. This rank-deficiency violates the fundamental assumption of 
MUSIC/ESPRIT \cite{huang1991near,zhi2007near,ramezani2025massive} that each steering vectors is orthogonal 
to the noise subspace. Therefore, conventional subspaced-based methods cannot be used without 
a prior user separation or decorrelation stage.

\subsection{Contributions of This Work}
In this paper, we investigate the problem of uplink joint multi-user channel estimation for 
near-field ELAA  mmWave/THz OFDM systems. To accommodate a large number of users that can be supported by
ELAA systems, we consider non-orthogonal uplink transmission scenarios where the length of pilot sequence is 
less than the number of served users. We model the received 
signal as a third-order low-rank tensor. Based on this model, we develop a tensor decomposition-based 
framework for channel parameter estimation. For the line-of-sight (LoS)-dominated scenario, the received signal is formulated
as a third-order tensor that
has a canonical polyadic decomposition (CPD) model, which is solved via an alternating least squares 
(ALS) algorithm. For the non-line-of-sight (NLoS) scenario, the formulated tensor follows a block term decomposition (BTD) model 
and the decomposition can be performed using a nonlinear least squares (NLS) algorithm. The channel parameters and user 
association are then extracted from the resulting factor matrices using the correlation-based 
methods. In addition, we provide uniqueness analysis for both LoS and NLoS cases, showing that the 
multi-user channel parameters can be jointly and uniquely recovered under the proposed tensor model 
{when the required k-rank conditions are satisfied. In particular, this condition may be met with a pilot length as small as $T\geq2$.}
By exploiting the intrinsic geometry of the LoS paths, user locations can be precisely estimated.
Simulations demonstrate that 
the proposed method significantly outperforms the CS-based baselines in channel estimation accuracy 
and achieves a high user localization precision.

\emph{Notations}: $\boldsymbol{a}$, $\boldsymbol{A}$ and $\mathcal{A}$ denote a vector, a matrix 
and a tensor, respectively; $(\cdot)^T$, $(\cdot)^H$, $(\cdot)^{-1}$ and $(\cdot)^{\dagger}$ denote 
the transpose, conjugate transpose, inverse and pseudo-inverse, respectively; $\boldsymbol{A}(:,m)$ 
and $\boldsymbol{A}(:,m:n)$ denote the $m$-th column of $\boldsymbol{A}$ and the submatrix of 
$\boldsymbol{A}$ from the $m$-th to the $n$-th columns, respectively; $\|\cdot\|_2$ and $\|\cdot\|_F$ 
denote the 2-norm and Frobenius norm, respectively; 
$\otimes$, $\odot$ and $\circ$ denote the Kronecker, Khatri-Rao and outer products, respectively; 
$\boldsymbol{I}_n$, $\boldsymbol{1}_{m}$, $\boldsymbol{0}_m$ and $\boldsymbol{0}_{m\times n}$ denote 
an $n\times n$ identity matrix, a $m \times 1$ all-ones vector, a $m \times 1$ all-zeros vector and 
a $m \times n$ all-zeros matrix, respectively; $\text{diag}(a_1,\cdots,a_M)$ and $\text{blkdiag}
(\boldsymbol{A}_1,\cdots,\boldsymbol{A}_M)$ denote a $M \times M$ diagonal matrix with $a_1,\cdots,a_M$ 
placed along its diagonal and a block-diagonal matrix with matrices $\boldsymbol{A}_1,\cdots,\boldsymbol{A}_M$ 
placed along its main diagonal, respectively; $\mathcal{U}(a,b)$ and $\mathcal{CN}(\mu,\sigma^2)$ 
denote a uniform distribution within range $(a,b)$ and a complex Gaussian distribution with mean 
$\mu$ and variance $\sigma^2$, respectively.

\section{System Model}
\begin{figure}
	\centering
	\includegraphics[width=0.9\linewidth]{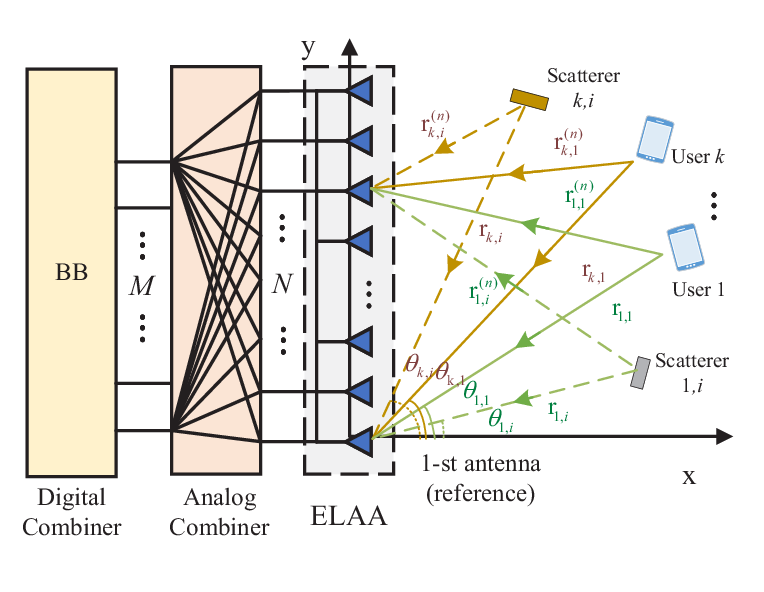}
	\caption{An ELAA system with multiple users in the near-field.}
	\label{fig_elaa}
\end{figure}

We consider the problem of uplink channel estimation for ELAA mmWave/THz systems, as shown in Fig. 
\ref{fig_elaa}, where the base station (BS), equipped with an extremely large-scale uniform linear 
array (ULA) with $N$ antenna elements, serves a number of single-antenna users.
We assume the system has a bandwidth of $B$ Hz 
and consists of $P$ sub-carriers. Let $f_c$ denote the center carrier 
frequency, and the corresponding wavelength is $\lambda_c = c/f_c$, where 
$c$ is the speed of light. The frequency associated with 
the $p$-th sub-carrier is expressed as $f_p = f_c + \frac{2p-P-1}{2(P-1)}B,p=1,\cdots,P$, 
with the corresponding wavelength given by $\lambda_p = c /f_p$.

The antenna spacing between two adjacent antennas at the BS is set to 
$d = \lambda_c /2$. Thus the array aperture and the Rayleigh 
distance are respectively given as $D = (N-1)d$ 
and $d_{R} = \frac{2D^2}{\lambda_c}$. To balance the hardware cost and efficiency in mmwave/THz, we consider
a hybrid analog and digital beamforming structure employed by the BS \cite{ning2021terahertz}, 
where the number of radio frequency (RF) chains $M$ is much less than the number of antenna elements  
but no smaller than the number of users, i.e., $K \leq M \ll N$.


\subsection{Channel Model}
As the number of antennas at the BS grows, the Rayleigh distance becomes 
comparable to the coverage radius. For example, for systems operating
at $30$ GHz, the Rayleigh distance $d_{R}$ is around $80.6$ meters for a half-wavelength ULA with 
$N=128$ antennas. This implies that users are highly likely located in the near-field region. 
In this case, the traditional planar wavefront model should be replaced by the more accurate 
spherical wavefront model. {Moreover, we assume that the system has a bandwidth that is 
much smaller than the carrier frequency, i.e., $B \ll f_c$. Under this assumption, the frequency-dependent variation of 
the array response across different subcarriers can be safely neglected.}
Specifically, the channel between the $k$-th user and the BS at the $p$-th 
sub-carrier can be characterized as
\begin{align}
	\boldsymbol{h}_{p,k} &= \sum\nolimits_{l=1}^{L_k} \alpha_{k,l}
	e^{-j2\pi f_p\tau_{k,l}}
	\boldsymbol{b}(\theta_{k,l},r_{k,l}) ,
	\label{h_channel}
\end{align}
where $L_k$ is the total number of signal paths between the BS and the 
$k$-th user; $\{\alpha_{l,k},\tau_{l,k},\theta_{l,k},r_{l,k}\}$ 
respectively denote the complex channel gain, the time delay, the angle and the 
distance from the reference antenna to the user/scatterer 
associated with the $l$-th path; and $\boldsymbol{b}(\theta,r)$ 
is the near-field steering vector given by
\begin{align}
	\boldsymbol{b}(\theta,r) = \frac{1}{\sqrt{N}}\bigl[
	e^{-j\frac{2\pi}{\lambda}(r^{(1)}-r)} \ \cdots 
	e^{-j\frac{2\pi}{\lambda}(r^{(N)}-r)} \bigl]^T,
	\label{b_steering}
\end{align}
where $r^{(n)}$ denotes the distance between the user/scatterer and 
the $n$-th antenna of the BS. For simplicity, we choose the first 
antenna as the reference antenna and we have $r^{(1)}\equiv r$. 
According to the geometric relationship, $r^{(n)}$ can be written as 
\begin{align}
	r^{(n)} 
	& = \sqrt{r^2 + (n-1)^2d^2-2r(n-1)d\sin\theta}
	\nonumber\\
	&\stackrel{(a)}{\approx} r - (n-1) d \sin\theta
	+ \frac{(n-1)^2d^2\cos^2\theta}{2r},
	\label{r_approx}
\end{align}
where $(a)$ follows from the widely used Fresnel approximation 
\cite{cui2022channel}. Thus, the near-field steering vector 
$\boldsymbol{b}(\theta,r)$ can be further simplified as 
\begin{align}
	\boldsymbol{b}(\theta,r) \approx \frac{1}{\sqrt{N}}
	[1 \ \cdots \ e^{-j\frac{2\pi}{\lambda}\left(-(N-1)d  \sin\theta + 
		\frac{(N-1)^2d^2\cos^2\theta}{2r}\right)} ]^T	 .
	\label{b_steering_approx}
\end{align}

\section{Problem Formulation}
In this paper, we consider the problem of uplink channel estimation and sensing based on the
received pilot signal at the BS. Here ``sensing'' specifically means the localization of users based on the estimated near-field channel parameters. Note that the
uplink channel estimation problem can be 
decomposed into a number of parallel single-user channel estimation 
problems if different users utilize mutually orthogonal pilot sequences. 
Nevertheless, in this work, we consider a joint multi-user tensor decomposition-based 
channel estimation scheme, which uses non-orthogonal pilot sequences to accommodate a growing number 
of users in ELAA systems while avoiding excessive pilot overhead.

\subsection{Received Signal Model}
The training sequence consists of $T$ consecutive OFDM symbols. 
At each time instant $t$, all users send their pilot vectors $\boldsymbol{s}_{k}[t] = 
[s_{1,k}[t] \ s_{2,k}[t] \ \cdots \ s_{P,k}[t])]^T \in 
\mathbb{C}^{P}, \forall k$ to the BS, where $s_{p,k}[t]$ 
denotes the $t$-th pilot symbol associated with the $k$-th user and the $p$-th 
sub-carrier. Suppose that the pilots are common for all sub-carriers, 
i.e., $s_{p,k}[t] = s_k[t],\forall p$. At the BS, each 
RF chain, i.e., the $m$-th RF chain, employs an individual combining vector 
$\boldsymbol{w}_m \in 
\mathbb{C}^{N}$ to combine the received antenna signals.
Considering the constant-modulus constraint imposed by analog beamforming, 
we assume that the BS adopts a random phase-only beamformer. Specifically,  
each entry of $\boldsymbol{w}_m$ is generated as $w_{mn} =e^{j\Phi_{mn}}$,$\forall m,n$, 
where $\Phi_{mn} \sim \mathcal{U}(0,2\pi)$ follows a uniform distribution.
After analog combining, the cyclic prefix is removed and symbols
are converted back to the frequency domain by performing
a discrete Fourier transform (DFT). As a result, the
received signal associated with the $p$-th sub-carrier and the 
$m$-th RF chain during the $t$-th time instant can be expressed as 
\begin{align}
	&y_{p,m}[t] = \boldsymbol{w}_{m}^H\sum_{k=1}^{K} 
	\boldsymbol{h}_{p,k} s_{k}[t] + \boldsymbol{w}_{m}^H
	\boldsymbol{n}_{p}[t] 
	\nonumber\\
	&= \boldsymbol{w}_m^H\sum_{k=1}^{K} \sum_{i=1}^{L_k}
	\alpha_{k,i} e^{-j2\pi f_p \tau_{k,i}}
	\boldsymbol{b}(\theta_{k,i},r_{k,i})
	s_k[t] + \tilde{n}_{p,m}[t],
	\label{y1}
\end{align}  
where $\boldsymbol{h}_{p,k}$ represents the channel between the BS and 
the $k$-th user at the $p$-th sub-carrier; and $\tilde{n}_{p,m}[t] 
\triangleq \boldsymbol{w}_{m}^H\boldsymbol{n}_{p}[t]$ represents the 
effective additive white Gaussian noise (AWGN). 
For notation convenience, all propagation paths from different users are 
indexed by a {global} path index $l$. Specifically, {
for the $i$-th path of the $k$-th user, its corresponding global index is 
$l = \sum_{j=1}^{k-1} L_j + i$. Under this mapping, we have}
$\alpha_l = \alpha_{k,i}$, $\theta_l = \theta_{k,i}$, $r_{l} = r_{k,i}$, 
and $\tau_l = \tau_{k,i}$. Moreover, the transmitted symbol associated 
with the $l$-th path is given by 
\begin{align}
	\tilde{s}_l[t] = s_k[t], \quad \forall l \in \left[{\sum\nolimits_{j=1}^{k-1}
	L_j+1,\sum\nolimits_{j=1}^{k}L_j}\right].
\end{align}

Thus, the received signal corresponding to the $p$-th subcarrier, the $m$-th RF chain and 
the $t$-th time instant can be re-written as 
\begin{align}
	{y}_{p,m}[t] & =  \boldsymbol{w}_m^H  \sum_{l=1}^{L} 
	\alpha_{l}e^{-j2\pi f_p \tau_{l}}\boldsymbol{b}(\theta_l,r_l)\tilde{{s}}_l[t]
	+ \tilde{n}_{p,m}[t]
	\nonumber\\
	& = \sum_{l=1}^{L} g[p](\tau_l) \times a[m](\alpha_l,\theta_l,r_l) \times \tilde{{s}}_l[t]+ \tilde{n}_{p,m}[t],
	\label{y4}
\end{align}
where $L = \sum_{k=1}^{K} L_k$ denotes the total number of paths 
associated with all users; $g[p](\tau_l) \triangleq e^{-j2\pi f_p \tau_{l}}$ denotes 
the frequency-domain phase response of the $l$-th path on the $p$-th subcarrier due to the time delay 
$\tau_l$; and $a[m](\alpha_l,\theta_l,r_l) \triangleq \alpha_l \boldsymbol{w}_m^H\boldsymbol{b}(\theta_l,r_l)$ 
represents the effective beamformed complex gain of the $l$-th path at the $m$-th RF chain, 
which jointly depends on the path gain $\alpha_l$, angle $\theta_l$, and distance $r_l$.



\subsection{Low-Rank Tensor Representation}
To reveal the inherent multi-linear structure of the received signals across all 
the frequencies,  RF chains, and time-instants, the received signal can be 
re-written as a third-order tensor $\mathcal{Y} \in \mathbb{C}^{P\times M \times T}$, 
whose $(p,m,t)$-th entry corresponds to $y_{p,m}[t]$. 
For the third-order tensor $\mathcal{Y}$, it is clear that the contribution from the $l$-th path 
can be represented as a rank-one tensor, which is the outer product of the following three vectors:
\begin{align}
	&\boldsymbol{g}(\tau_l) \triangleq \left[g[1](\tau_l) \ 
	\cdots \ g[P](\tau_l)\right]^T \in \mathbb{C}^{P}, \nonumber\\
	&\boldsymbol{a}(\alpha_l,\theta_l,r_l) \triangleq \left[
	a[1](\alpha_l,\theta_l,r_l) \ \cdots \ a[M](\alpha_l,\theta_l,r_l)\right]^T
	\in \mathbb{C}^{M}, \nonumber\\
	&\tilde{\boldsymbol{s}}_l \triangleq  \big[\tilde{s}_{l}[1] \ \cdots \ 
	\tilde{s}_{l}[T]\big]^T \in \mathbb{C}^{T}.
	\label{r1_component}
\end{align}
Consequently, the received tensor $\mathcal{Y}$ admits a decomposition into a 
sum of rank-one component tensors, i.e.,
\begin{align}
	\mathcal{Y} = \sum_{l=1}^{L} \boldsymbol{g}(\tau_l) \circ 
	\boldsymbol{a}(\alpha_l,\theta_l,r_l) \circ \tilde{\boldsymbol{s}}_l+ 
	\mathcal{N},
	\label{CPD}
\end{align}
where $\mathcal{N} \in \mathbb{C}^{P\times M \times T}$ denotes 
the noise tensor, whose $(p,m,t)$-th entry corresponds to $\tilde{n}_{p,m}[t]$. 
The tensor decomposition in (\ref{CPD}) is known as the CPD. By collecting 
the factor vectors associated with all 
$L$ propagation paths, we obtain the factor matrices associated with the noiseless part 
of $\mathcal{Y}$:
\begin{align}
	\boldsymbol{G} &\triangleq [\boldsymbol{g}(\tau_1) \ \cdots \ 
	\boldsymbol{g}(\tau_L)] \in \mathbb{C}^{P\times L},
	\nonumber\\
	\boldsymbol{A} &\triangleq [\boldsymbol{a}(\alpha_1,\theta_1,r_1) \ 
	\cdots \ \boldsymbol{a}(\alpha_L,\theta_L,r_L)] \in \mathbb{C}^{M \times L},
	\nonumber\\
	\boldsymbol{S} &\triangleq [\tilde{\boldsymbol{s}}_1 \ \cdots \ \tilde{\boldsymbol{s}}_L] 
	\in \mathbb{C}^{T\times L}.
	\label{mtx_fac}
\end{align}
Here, the factor matrix $\boldsymbol{G}$ is characterized by different propagation delays across 
all subcarriers, $\boldsymbol{A}$ captures complex path losses and
near-field spatial responses associated with different paths, and 
$\boldsymbol{S}$ is a matrix with its column corresponding to a certain user's pilot sequence.

The tensor form (\ref{CPD}) of the received signals allows us to find an efficient way to estimate the wireless channel. 
Under certain mild uniqueness conditions, the received tensor $\mathcal{Y}$ admits an essentially unique decomposition, up to scaling and permutation ambiguities. Thus we can first estimate the three factor matrices and then extract the associated channel parameters $\left\{\tau_l,\theta_l,r_l,\alpha_l\right\}_{l=1}^L$ from the recovered factor matrices. Specifically, the exact tensor decomposition model and the corresponding uniqueness conditions depend on the underlying channel propagation structure. In view of this, we discuss channel estimation for the LoS-dominated and NLoS scenarios in Sections~IV and~V, respectively.


\section{Channel Estimation for LoS-dominated Scenarios}\label{pro_single}
In this section, we first consider the scenario where the LoS path between the BS
and each user is not blocked. In mmWave/THz communication systems, the energy of NLoS paths is typically 
much smaller than that of the LoS path \footnote{Many channel measurement campaigns reveal that the power 
of the mmWave/THz LoS path is much higher (about 13 dB higher over the mmWave band and 20 dB higher over 
the THz band) than the sum of power of NLoS paths \cite{akdeniz2014millimeter}}, making their contribution 
to the received signal negligible. Therefore, only the LoS path is modeled in this work for simplicity.
In this case, we have {$L_k = 1$ and $L = K$}, and the received signal can be expressed as
\begin{align}
	\mathcal{Y} = \sum\nolimits_{k=1}^{K}\boldsymbol{g}(\tau_k) \circ 
	\boldsymbol{a}(\alpha_k,\theta_k,r_k) \circ \boldsymbol{s}_k+ 
	\mathcal{N},
	\label{CPD_los}
\end{align}
which admits a CPD with the following factor matrices:
\begin{align}
	\boldsymbol{G}_o &\triangleq [\boldsymbol{g}(\tau_1) \ \cdots \ 
	\boldsymbol{g}(\tau_K)] \in \mathbb{C}^{P\times K},
	\nonumber\\
	\boldsymbol{A}_o &\triangleq [\boldsymbol{a}(\alpha_1,\theta_1,r_1) \ 
	\cdots \ \boldsymbol{a}(\alpha_K,\theta_K,r_K)] \in \mathbb{C}^{M \times K},
	\nonumber\\
	\boldsymbol{S}_o &\triangleq [\boldsymbol{s}_1 \ 
	\cdots \ \boldsymbol{s}_K] 
	\in \mathbb{C}^{T\times K}.
	\label{mtx_fac_los}
\end{align}
where $\boldsymbol{s}_k \triangleq \left[s_k[1] \ \cdots \ s_k[T]\right]^T$.

\subsection{Uniqueness Condition}
Before proceeding to the CPD, we first discuss under what conditions the 
uniqueness of the CPD can be guaranteed. A well-known sufficient condition is 
Kruskal's condition \cite{kruskal1977three}, which is stated in Theorem 
\ref{thero_kkc}. 

\newtheorem{theorem}{Theorem}
\begin{theorem}
	\label{thero_kkc}
	Let $k_A$ denote the k-rank of $\boldsymbol{A}$, which 
	is defined as the maximum number $k_A$ such that any $k_A$ columns 
	of $\boldsymbol{A}$ are linearly independent.
	Let $(\boldsymbol{A},\boldsymbol{B},\boldsymbol{C})$ be a CP solution which 
	decomposes a third-order tensor $\mathcal{Y} \in \mathbb{C}^{I_1\times I_2 \times I_3}$
	into $Q$ rank-one components, where $\boldsymbol{A} \in \mathbb{C}^{I_1\times Q}$, 
	$\boldsymbol{B} \in \mathbb{C}^{I_2\times Q}$, $\boldsymbol{C} \in \mathbb{C}^
	{I_3\times Q}$. If the Kruskal condition
	\begin{align}
		k_{A} + k_{B} + k_{C} \geq 2Q+2 
	\end{align}
	is satisfied, then the CPD of $\mathcal{Y}$ is unique up to scaling and 
	permutation ambiguities. Specifically, the scaling and permutation ambiguities may 
	lead to an alternative CP solution 
	$(\bar{\boldsymbol{A}},\bar{\boldsymbol{B}},\bar{\boldsymbol{C}})$, which also 
	decomposes $\mathcal{Y}$ into $Q$ rank-one components and it satisfies 
	$\bar{\boldsymbol{A}} = \boldsymbol{A}\boldsymbol{\Lambda}_A\boldsymbol{\Pi}$,
	$\bar{\boldsymbol{B}} = \boldsymbol{B}\boldsymbol{\Lambda}_B\boldsymbol{\Pi}$,
	and $\bar{\boldsymbol{C}} = \boldsymbol{C}\boldsymbol{\Lambda}_C\boldsymbol{\Pi}$, 
	where $\boldsymbol{\Lambda}_A$, 
	$\boldsymbol{\Lambda}_B$ and $\boldsymbol{\Lambda}_C$ are unique diagonal matrices 
	such that $\boldsymbol{\Lambda}_A\boldsymbol{\Lambda}_B\boldsymbol{\Lambda}_C = 
	\boldsymbol{I}_{Q}$, and $\boldsymbol{\Pi}$ is a unique permutation matrix. 
\end{theorem}

From the above theorem, the CPD of $\mathcal{Y}$ is essentially unique if the following Kruskal 
condition is satisfied
\begin{align}
	k_{G_o} + k_{A_o} + k_{S_o} \geq 2K + 2.\label{krus1}
\end{align}
We next examine the k-ranks of the involved factor matrices. 

We first consider the k-rank of $\boldsymbol{G}_o$. Since $\boldsymbol{G}_o$ 
is a columnwise-scaled Vandermonte matrix characterized by distinct delay parameters $\{\tau_l\}$, the k-rank of $\boldsymbol{G}_o$ is 
equal to the smallest value of $P$ and $K$, i.e.,
\begin{align}
	k_{G_o} = \min (P,K).
\end{align}
In practical OFDM systems, it is reasonable to assume that the number of 
sub-carriers allocated for training is greater than the number of users, 
i.e. $P \geq K$. Under this condition, we have $k_{G_o} = K$.


We now turn to the k-rank of $\boldsymbol{A}_o$. Recall that the factor matrix $\boldsymbol{A}_o$ 
is obtained by projecting the weighted near-field steering vectors onto a reduced-dimensional 
RF domain through the analog combining matrix  $\boldsymbol{W}$, i.e., $\boldsymbol{A}_o \triangleq 
\boldsymbol{W}^H \boldsymbol{B} \boldsymbol{D}_{\alpha}$,
where 
\begin{align}
	&\boldsymbol{W} \triangleq \big[\boldsymbol{w}_{1} \ \cdots \ 
	\boldsymbol{w}_{M}\big] \in \mathbb{C}^{N\times M},
	\nonumber\\
	&\boldsymbol{D}_{\alpha} \triangleq \text{diag}\{\alpha_1,\cdots,\alpha_K\}
	\in \mathbb{C}^{K\times K},
	\nonumber\\
	&\boldsymbol{B} \triangleq \left[
	\boldsymbol{b}(\theta_1,r_1) \ \cdots \ \boldsymbol{b}(\theta_K,r_K)\right] 
	\in \mathbb{C}^{N\times K},
\end{align}
Generally, steering vectors with distinct angle-distance pairs are linearly 
independent with high probability 
\cite{cui2022channel}. Consequently, the near-field array response matrix $\boldsymbol{B}$
is full column rank when $K \leq N$. 
Note that $\boldsymbol{D}_{\alpha}$ is a non-singular diagonal matrix. Hence $\boldsymbol{B}_{\alpha}\triangleq\boldsymbol{B}\boldsymbol{D}_{\alpha}$
is full column rank.
As a result, the k-rank of $\boldsymbol{A}_o$ depends on whether the RF-domain projection 
$\boldsymbol{W}^H$ preserves the rank of $\boldsymbol{B}_{\alpha}$.

Recall that a random phase-only combining matrix $\boldsymbol{W}$ is employed, under which 
$\boldsymbol{W}$ is full column rank with probability one.
For any subset of $k_{A_o}$ columns from $\boldsymbol{B}_{\alpha}$ 
(where $k_{A_o} \le K$), their linear independence is preserved after projection if none of the 
linear combinations of these $k_{A_o}$ columns lies in the null space of $\boldsymbol{W}^H$. 
Since entries of $\boldsymbol{W}^H$ are randomly generated, the probability that the 
linear combination of these $k_{A_o}$ columns lies in the null space of $\boldsymbol{W}^H$ is zero, provided that $k_{A_o} \le M$. 
As a result, with probability one, any subset of at most $\min\{M,K\}$ columns of 
$\boldsymbol{A}_o$ is linearly independent, which implies that the k-rank of $\boldsymbol{A}_o$ 
satisfies
\begin{align}
	k_{A_o} = \min\{M,K\} = K.
\end{align}

Based on the above results, we know that the Kruskal condition in (\ref{krus1}) can  
be satisfied by designing
a set of training sequences $\boldsymbol{S}_o$ such that 
$k_{S_o} \geq 2$, namely, any two columns of $\boldsymbol{S}_o$ are linearly 
independent. This condition can be readily met when $T \geq 2$ and the pilot vectors 
of different users are linearly independent. To 
be concrete, we can design the training pilots by minimizing the mutual 
coherence of $\boldsymbol{S}_o$, i.e.,
\begin{align}
	\min_{\boldsymbol{S}_o} \max_{i\neq j} \frac{
		|\boldsymbol{s}_i^H\boldsymbol{s}_j|}{\|\boldsymbol{s}_i\|_2
		\|\boldsymbol{s}_j\|_2},
	\label{s_design}
\end{align}
where $\boldsymbol{s}_i$ denotes the training pilot vector associated 
with the $i$-th user. This optimization is fundamentally a Grassmannian 
line packing problem \cite{conway1996packing}, which aims to find a set of 
lines in a complex space that are as far apart as possible.

{The aforementioned analysis reveals that the proposed CPD framework can ensure
the essential uniqueness of the channel decomposition even in the underdetermined 
regime where $T\ll K$. Specifically, the pilot length can be substantially reduced 
to $T =2$ when both $k_{G_o}$ and $k_{A_o}$ are equal to $K$, which is usually 
the case when angular and distance parameters associated with different users are 
sufficiently separated.}

{By further exploiting the Vandermonde structure of the factor matrix 
$\boldsymbol{G}_o$, the uniqueness condition can be relaxed through the spatial 
smoothing framework proposed in \cite[Theorem III.3]{Sorensen2013blind}. 
Specifically, the considered Vandermonde-constrained CPD remains essentially unique if
\begin{align}
	\left\lceil \frac{K}{M} \right\rceil+
	\left\lceil \frac{K}{T} \right\rceil
	\le P. \label{unq2_vd}
\end{align}
Compared with the classical uniqueness condition in \eqref{krus1}, the 
above Vandermonde-constrained condition is less restrictive and applies to a 
broader range of $(P,M,T)$ tuples.}

{Overall, the above uniqueness results demonstrate that the proposed
tensor model remains identifiable even under non-orthogonal pilot transmission. 
They provide the theoretical foundation for the subsequent CPD-based integrated 
channel estimation and sensing method.}

\subsection{CP Decomposition}
The CPD can be accomplished by solving the following optimization 
problem 
\begin{align}
	\min_{\boldsymbol{G}_o,\boldsymbol{A}_o,\boldsymbol{S}_o} \quad 
	\left\|\mathcal{Y} - \sum\nolimits_{k=1}^{K}\hat{\boldsymbol{g}}_k \circ 
	\hat{\boldsymbol{a}}_k \circ  \hat{\boldsymbol{s}}_{k}\right\|_F^2.
	\label{p_cp1}
\end{align}
where $\hat{\boldsymbol{g}}_k$, $\hat{\boldsymbol{a}}_k$ and $\hat{\boldsymbol{s}}_{k}$ 
denote the $k$-th column of $\boldsymbol{G}_o$, $\boldsymbol{A}_o$ and 
$\boldsymbol{S}_o$, respectively.
{Since uplink channel estimation is typically performed after a series of 
handshaking procedures, the number of users, $K$, can be assumed to be known 
\emph{a priori}.} The above optimization can be efficiently solved by an ALS 
procedure, which iteratively minimizes the data fitting error with respect to 
(w.r.t.) the three factor matrices:
\begin{align}
	\boldsymbol{G}_o^{(t+1)} &= \arg \min_{\boldsymbol{G}_o} \left\|\boldsymbol{Y}_{(1)}^T
	- (\boldsymbol{S}_o^{(t)}\odot \boldsymbol{A}_o^{(t)})\boldsymbol{G}_o^{T}\right\|_F^2,
	\\
	\boldsymbol{A}_o^{(t+1)} &= \arg \min_{\boldsymbol{A}_o} \left\|\boldsymbol{Y}_{(2)}^T
	- (\boldsymbol{S}_o^{(t)}\odot \boldsymbol{G}_o^{(t+1)})\boldsymbol{A}_o^{T}\right\|_F^2,
	\\
	\boldsymbol{S}_o^{(t+1)} &= \arg \min_{\boldsymbol{S}_o} \left\|\boldsymbol{Y}_{(3)}^T- (\boldsymbol{A}_o^{(t+1)}\odot 
	\boldsymbol{G}_o^{(t+1)})\boldsymbol{S}_o^{T}\right\|_F^2,
	\label{p_cp2_s}
\end{align}
where $\boldsymbol{Y}_{(n)}, n= 1,2,3$ denotes the mode-$n$
unfolding of the tensor $\mathcal{Y}$.

{At the beginning of the ALS algorithm, the factor matrices are initialized using a 
pilot-aided truncated singular value decomposition (SVD) scheme. Specifically, 
the received tensor is first unfolded along the pilot mode
\begin{align}
	\boldsymbol{Y}_{(3)} = (\boldsymbol{G}_o \odot \boldsymbol{A}_o)\boldsymbol{S}_o^T 
	+ \boldsymbol{N}_{(3)}\in 
	\mathbb{C}^{MP\times T}.
\end{align}
Since the pilot matrix $\boldsymbol{S}_o$ is known a \emph{priori} at the BS, its 
mixing effect can be mitigated through a pseudo-inverse operation, i.e., 
$\boldsymbol{X}_{(3)} = \boldsymbol{Y}_{(3)} \boldsymbol{S}_o^{*}
(\boldsymbol{S}_o^T\boldsymbol{S}_o^{*} + \epsilon\boldsymbol{I}_{K})^{-1}  \in 
\mathbb{C}^{MP \times K}$, where $\epsilon > 0$ denotes the regularization parameter. 
Then, the $k$-th column of $\boldsymbol{X}_{(3)}$ can be reshaped into a $P \times M$ 
matrix $\boldsymbol{X}_k$. By performing the singular value decomposition (SVD) of 
$\boldsymbol{X}_k$, the initial factor matrices are obtained as 
\begin{align}
	\hat{\boldsymbol{g}}_k^{(0)} = \boldsymbol{v}_{k,\max}\lambda_{k,\max}^{1/2}, \ \
	\hat{\boldsymbol{a}}_k^{(0)} = \boldsymbol{u}^{*}_{k,\max}\lambda_{k,\max}^{1/2}, \ \
	\hat{\boldsymbol{s}}_k^{(0)} = \boldsymbol{s}_{k}.
	\label{cpd_ini}
\end{align}
where $\lambda_{k,\max}$ is the largest singular value of $\boldsymbol{X}_k$, 
and $\boldsymbol{v}_{k,\max}$ and $\boldsymbol{u}_{k,\max}$ are the corresponding left 
and right singular vectors, respectively.
This initialization provides a good starting point for the CPD
and can significantly accelerate the convergence of the subsequent ALS iterations.}

{In addition to ALS, we would like to point out that, by exploiting the Vandermonde 
structure of the factor matrix, a spatial-smoothing-aided Vandermonde-constrained CPD 
algorithm can also be employed to solve the problem \eqref{p_cp1} \cite{Sorensen2013blind}, 
which can provide an analytical solution without involving a cumbersome iterative process.}

As discussed earlier, the CPD is unique up to scaling and permutation 
ambiguities under some mild conditions. Specifically, the relationship between 
the estimated factor matrices and the true factor matrices is given by 
\begin{align}
	\hat{\boldsymbol{G}}_o &= \boldsymbol{G}_o\boldsymbol{\Lambda}_1\boldsymbol{\Pi}
	+\boldsymbol{E}_1,
	\nonumber\\
	\hat{\boldsymbol{A}}_o &= \boldsymbol{A}_o\boldsymbol{\Lambda}_2\boldsymbol{\Pi}
	+\boldsymbol{E}_2,
	\nonumber\\
	\hat{\boldsymbol{S}}_o &= \boldsymbol{S}_o\boldsymbol{\Lambda}_3\boldsymbol{\Pi}
	+\boldsymbol{E}_3,
\end{align}
where $\boldsymbol{\Lambda}_1$, $\boldsymbol{\Lambda}_2$ and $\boldsymbol{\Lambda}_3$ 
are nonsingular diagonal matrices and satisfy $\boldsymbol{\Lambda}_1\boldsymbol{\Lambda}_2
\boldsymbol{\Lambda}_3 =\boldsymbol{I}$; $\boldsymbol{\Pi}$ is a permutation 
matrix; $\boldsymbol{E}_1$, $\boldsymbol{E}_2$ and $\boldsymbol{E}_3$ denote the 
estimation errors associated with the three estimated factor matrices, 
respectively. 

\subsection{Channel Estimation}
We now discuss how to extract the channel parameters from the estimated factor matrices. Ignoring the estimation errors $\boldsymbol{E}_3$, it can be observed 
that each column of $\hat{\boldsymbol{S}}_{o}$ is a scaled version of 
a training sequence associated with a certain user. Note that all
available training pilots are known \emph{a prior} at the BS. 
Hence, the permutation matrix $\boldsymbol{\Pi}$ can be estimated via 
a simple correlation-based method. More precisely, the user 
index associated with the $k$-th column of $\hat{\boldsymbol{S}}_o$ can 
be estimated as 
\begin{align}
	\hat{i}_{k} = \arg \max_{i\in\{1,\cdots,K\}} \frac{|\boldsymbol{s}_i^H\hat{\boldsymbol{s}}_k|}
	{\|\boldsymbol{s}_i\|_2\|\hat{\boldsymbol{s}}_k\|_2}, \ 
	\forall k = 1,\cdots,K.
	\label{ue_match_s}
\end{align}
If more than two columns choose the same user index, there 
exists a conflict. The column with a higher correlation retains this index, 
while the other column chooses a new index from the remaining 
indices by repeating (\ref{ue_match_s}). After determining the user index
associated with each column, the permutation matrix $\boldsymbol{\Pi}$ can 
be obtained as 
\begin{align}
	\boldsymbol{\Pi}(i,j) = \left\{\begin{array}{rcl}
		1,&(i,j)\in\{({\hat{i}_k},k)\}_{k=1}^{K},\\
		0,&\text{otherwise}.
	\end{array}\right.
\end{align}

Given $\boldsymbol{\Pi}$, the permutation ambiguity for the estimated factor 
matrices can be removed. Thus, we have 
\begin{align}
	\hat{\boldsymbol{G}}_o &= \hat{\boldsymbol{G}}_o\boldsymbol{\Pi}^T
	= \boldsymbol{G}_o\boldsymbol{\Lambda}_1 + \boldsymbol{E}_1,
	\nonumber\\
	\hat{\boldsymbol{A}}_o &=\hat{\boldsymbol{A}}_o\boldsymbol{\Pi}^T
	= \boldsymbol{A}_o\boldsymbol{\Lambda}_2 + 
	\boldsymbol{E}_2,
	\nonumber\\
	\hat{\boldsymbol{S}}_o &=\hat{\boldsymbol{S}}_o\boldsymbol{\Pi}^T
	= \boldsymbol{S}_o\boldsymbol{\Lambda}_3 + 
	\boldsymbol{E}_3.
	\label{est2_factor_s}
\end{align}

We can see that the $k$-th column of $\hat{\boldsymbol{G}}_o$ is a 
scaled version of $\boldsymbol{g}(\tau_k)$. Therefore, the delay parameter
$\tau_k$ associated with the $k$-th user can be estimated by maximizing the 
normalized correlation
\begin{align}
	\hat{\tau}_k = \arg \max_{\tau}\frac{\big|\boldsymbol{g}^H
		(\tau)\hat{\boldsymbol{g}}_k\big|}
	{\|\boldsymbol{g}(\tau)\|_2\|\hat{\boldsymbol{g}}_k\|_2}
	\label{est_t_s}
\end{align}
{To balance estimation precision and computational complexity, a 
multi-level coarse-fine search strategy is adopted. Specifically, the 
first search level is performed over a sufficiently large feasible 
parameter range to avoid missing the global optimum. The subsequent 
refinement levels progressively restrict the search region to the 
neighborhood of the optimal grid point obtained at the previous level, 
where a denser grid is generated for further refinement.}

Similarly, the $k$-th column of $\hat{\boldsymbol{A}}_o$ is a scaled version 
of $\tilde{\boldsymbol{b}}(\theta_k,r_k)$. {Therefore, the angle-distance 
pair can be estimated using the same multi-level coarse-fine search 
framework described above, where the search grid is constructed according to 
near-field polar-domain codebook sampling principle \cite{cui2022channel}. }
\begin{align}
	\hat{\theta}_k,\hat{r}_k = \arg \max_{\theta,r}  \frac{\big|
		\boldsymbol{b}^H(\theta,r)\boldsymbol{W}
		\hat{\boldsymbol{a}}_k\big|}
	{\|\boldsymbol{W}^H\boldsymbol{b}(\theta,r)\|_2
		\|\hat{\boldsymbol{a}}_k\|_2}.
	\label{est_thetar1}
\end{align} 

Moreover, in the LoS scenario, the time delay $\tau_k$ and the distance $r_k$ 
satisfy a deterministic relationship, i.e. $\tau_k = r_k / c$. Hence, the 
distance from the BS to the $k$-th user can be directly obtained as 
$\hat{r}_k = \hat{\tau}_k c$. With $\hat{r}_k$ readily determined, only 
the angle $\theta_k$ needs to be estimated through a {multi-level} 
coarse-fine search, i.e. 
\begin{align}
	{\hat{\theta}_k} = \arg \max_{\theta}  \frac{\big|
		\boldsymbol{b}^H(\theta,\hat{r}_k)\boldsymbol{W}
		\hat{\boldsymbol{a}}_k\big|}
	{\|\boldsymbol{W}^H\boldsymbol{b}(\theta,\hat{r}_k)\|_2
		\|\hat{\boldsymbol{a}}_k\|_2}.
	\label{est_theta}
\end{align}
Leveraging this prior relationship significantly reduces the computational 
complexity and improves the estimation accuracy, as it eliminates the joint 
range-angle search.

Next, we try to recover the path loss gains $\{\hat{\alpha}_k\}$. After 
obtaining $\{\hat{\tau}_k,\hat{\theta}_k,\hat{r}_k\}$, we can reconstruct 
\begin{align}
	\tilde{\boldsymbol{G}}_o &= [\boldsymbol{g}(\hat{\tau}_1) \ \cdots
	\ \boldsymbol{g}(\hat{\tau}_{K})],
	\nonumber\\
	\tilde{\boldsymbol{B}}_o &= \boldsymbol{W}^H [\boldsymbol{b}(
	\hat{\theta}_1,\hat{r}_1) \ \cdots \ \boldsymbol{b}(\hat{\theta}_{K},
	\hat{r}_{K})].
\end{align} 
Ignoring the estimation error, the scaling ambiguity matrix can be 
estimated as $\boldsymbol{\Lambda}_1 = \tilde{\boldsymbol{G}}_o^{\dagger}
\hat{\boldsymbol{G}}_o$ and ${\boldsymbol{\Lambda}_3 = \boldsymbol{S}_o^{\dagger}\hat{\boldsymbol{S}}_o}$. Moreover, since 
$\boldsymbol{\Lambda}_1\boldsymbol{\Lambda}_2\boldsymbol{\Lambda}_3 
= \boldsymbol{I}$, we have $\boldsymbol{\Lambda}_2^{\dagger} = 
\boldsymbol{\Lambda}_1\boldsymbol{\Lambda}_3$. 
Ideally, there should be $\hat{\boldsymbol{A}}_o = \tilde{\boldsymbol{B}}_o
\boldsymbol{D}_{\alpha} \boldsymbol{\Lambda}_2$. Hence, 
$\boldsymbol{D}_{a}$ can be estimated as 
\begin{align}
	\hat{\boldsymbol{D}}_{\alpha} = \text{diag}\left(
	\tilde{\boldsymbol{B}}_o^{\dagger}\hat{\boldsymbol{A}}_o\boldsymbol{\Lambda}_1
	\boldsymbol{\Lambda}_3\right).
	\label{est_alpha_s}
\end{align}

Finally, the channels $\{\boldsymbol{h}_{p,k}\}$ can be reconstructed 
by the estimated parameters $\{\hat{\tau}_k,\hat{\theta}_k,\hat{r}_k,
\hat{\alpha}_k\}$. With the obtained distance and angle information, 
the position of the $k$-th user is estimated as 
\begin{align}
	\hat{\boldsymbol{p}}_k = [\hat{r}_k\cos\hat{\theta}_k,
	\hat{r}_k\sin\hat{\theta}_k]^T.
\end{align}

{ For clarity, the overall processing flow of the proposed CPD-based framework is 
summarized as Algorithm \ref{alg_cpd}.}
\begin{algorithm}[H]
	\caption{CPD-based Integrated Channel Estimation and Sensing Framework}
	\begin{algorithmic}[1]
		\REQUIRE{$\mathcal{Y}$, $K$, $\boldsymbol{S}_o$, $\boldsymbol{W}$.}
		\ENSURE{$\{\hat{\boldsymbol{h}}_{p,k}\}_{k=1,p=1}^{K,P}$, $\{\hat{\boldsymbol{p}}_k\}_{k=1}^{K}$.}
		\STATE Initialize the factor matrices according to \eqref{cpd_ini}.
		\STATE Perform CPD via the ALS algorithm.
		\STATE Remove the permutation ambiguity of the recovered factor matrices according 
		to \eqref{ue_match_s}.
		\STATE Extract the channel parameters $\{\hat{\tau}_k,\hat{\theta}_k,\hat{r}_k\}$
		using the multi-level coarse-fine search in \eqref{est_t_s}--\eqref{est_theta}.
		\STATE Estimate the path gains $\{\hat{\alpha}_k\}$ according to \eqref{est_alpha_s}.
		\STATE Reconstruct the multi-user channels and estimate the user positions from the recovered angle-distance parameters.
	\end{algorithmic}
	\label{alg_cpd}
\end{algorithm}

\section{Channel Estimation for NLoS Scenarios}\label{pro_mul}
In this section, we discuss how to estimate the wireless channel for the general geometric multi-path scenario where
the LoS path is blocked and there exists multiple NLoS paths between the BS 
and each user. 
\subsection{CPD Model and Uniqueness Analysis}
In the NLoS scenario, the received signal still follows the general CPD model 
introduced in Eq. (\ref{CPD}), where each propagation path contributes a rank-one component 
to the received signal tensor. Therefore, the noiseless part of received signal can be compactly 
characterized by the factor matrices defined in (\ref{mtx_fac}). Note that in $\boldsymbol{S}$, 
columns associated with the same user are identical. 
Therefore, the factor matrices $\boldsymbol{S}$ can be further factorized as 
\begin{align}
	\boldsymbol{S} \triangleq \boldsymbol{S}_{o}\boldsymbol{O},
	\label{map_s}
\end{align}
where $\boldsymbol{S}_o$ and $\boldsymbol{s}_k$ are defined in (\ref{mtx_fac_los})
and $\boldsymbol{O}\in \mathbb{B}^{K\times L}$ 
is a binary mapping matrix that associates each path with its corresponding 
user. Specifically, the $(k,l)$-th entry of $\boldsymbol{O}$ is given by 
\begin{align}
	o_{kl} = \left\{
	\begin{array}{rcl}
		1, &\text{$l$-th path is associated with the $k$-th user},\\
		0, &\text{otherwise}.
	\end{array}\right.
\end{align}
As a result, the $k$-th row of $\boldsymbol{O}$ contains exactly $L_k$ 
nonzero entries. 

We next discuss the uniqueness condition of the CPD model in (\ref{CPD}) for NLoS scenarios. 
{Based} on the theorem \ref{thero_kkc}, Kruskal's condition for the general multi-path cases becomes 
\begin{align}
	k_G + k_B + k_S \geq 2L+2.
\end{align}
Since both the factor matrices $\boldsymbol{G} \in \mathbb{C}^{P\times L}$ 
and $\boldsymbol{A} \in \mathbb{C}^{M\times L}$ have a k-rank 
bounded by $L$, satisfying the above Kruskal condition strictly requires $k_S \geq 2$.
However, this is unattainable in multi-path scenarios, where the path components associated 
with a common user share the same pilot sequence, rendering the k-rank of $\boldsymbol{S}$ equal
to one. 
As a result, Kruskal's condition under a CPD model can never be satisfied 
for the general multi-path scenario. This implies that if we still treat the received 
signal as a CPD model, the decomposition is not unique. The estimated factor matrices 
will suffer from severe rotational ambiguities, making the channel parameters unidentifiable. 

\subsection{BTD Model and Uniqueness Analysis}\label{sec:nlos_unique}
To resolve this issue, we aggregate the collinear components associated with the 
same user into a block term, rather than treat them as $L_k$ independent rank-one path 
components. 
As such, the received signal in (\ref{CPD}) is reformulated as a BTD structure, 
\begin{align}
	\mathcal{Y} &= \sum\nolimits_{k = 1}^{K} \sum_{i=1}^{L_k} 
	\boldsymbol{g}(\tau_{k,i}) \circ 
	\boldsymbol{a}(\alpha_{k,i},\theta_{k,i},r_{k,i}) 
	\circ \boldsymbol{s}_{k} + 
	\mathcal{N}
	\nonumber\\
	&= \sum\nolimits_{k=1}^{K} (\boldsymbol{G}_k\boldsymbol{A}_k^T)
	\circ \boldsymbol{s}_{k} + 
	\mathcal{N},
	\label{y_tensor3}
\end{align}
where
\begin{align}
	\boldsymbol{G}_{k} &\triangleq \left[\boldsymbol{g}(\tau_{k,1}) \ \cdots \ 
	\boldsymbol{g}(\tau_{k,L_{k}})\right],
	\nonumber\\
	\boldsymbol{A}_{k} &\triangleq \left[\boldsymbol{a}(\alpha_{k,1},\theta_{k,1},r_{k,1}) \ \cdots \ 
	\boldsymbol{a}(\alpha_{k,L_k},\theta_{k,L_k},r_{k,L_k})\right],
	\label{block_ga1}
\end{align}
and $\boldsymbol{s}_k$ is defined after (\ref{mtx_fac_los}). In this formulation, the 
tensor $\mathcal{Y}$ is expressed as a sum of matrix-vector outer products, 
more specifically, a sum of rank-$(L_k,L_k,1)$ terms since 
$\boldsymbol{G}_k$ and $\boldsymbol{A}_k$ are both rank-$L_k$. 
The generalized Kruskal's condition provides sufficient conditions 
for the uniqueness of the above BTD
\cite{de2008decompositions}, which 
can be elaborated as follows. 

\begin{theorem}
	\label{thero_gkkc}
	Let $(\boldsymbol{A},\boldsymbol{B},\boldsymbol{C})$ be a BTD solution 
	which decomposes a third-order tensor $\mathcal{Y} \in \mathbb{C}^{I_1\times I_2 \times I_3}$
	into $Q$ rank-$(L_q,L_q,1)$ terms, i.e. 
	\begin{align}
		\mathcal{Y} = \sum\nolimits_{q=1}^{Q}(\boldsymbol{A}_q\boldsymbol{B}_q^T)
		\circ \boldsymbol{c}_q ,
	\end{align}
	where $\boldsymbol{A}_q\in \mathbb{C}^{I_1\times L_q}$, $\boldsymbol{B}_q
	\in \mathbb{C}^{I_2\times L_q}$ and $\boldsymbol{c}_q\in \mathbb{C}^{I_3}$ 
	are the factor matrices/vectors associated with the $q$-th block term. 
	Here we define $\boldsymbol{A} = \left[\boldsymbol{A}_1 \ \cdots \ \boldsymbol{A}_Q
	\right]\in \mathbb{C}^{I_1\times L}$, $\boldsymbol{B} = \left[\boldsymbol{B}_1 
	\ \cdots \ \boldsymbol{B}_Q\right]\in \mathbb{C}^{I_2\times L}$ and $\boldsymbol{C} = 
	\left[\boldsymbol{c}_1 \ \cdots \ \boldsymbol{c}_Q\right]\mathbb{C}^{I_3\times Q}$
	with $L = \sum_{q=1}^{Q}L_q$. Moreover, we assume $I_1 \geq \max_{q} L_q$, 
	$I_2 \geq\max_{q} L_q$, 
	$\text{rank}(\boldsymbol{A}_q) = L_q$ and $\text{rank}(\boldsymbol{B}_q) = L_q$. 
	If the following conditions 
	\begin{align}
		I_1I_2 \geq \sum\nolimits_{q=1}^{Q}L_q^2,
		\\
		k^{'}_{A} + k^{'}_{B} + k_{C} \geq 2Q+2 ,
	\end{align}
	are satisfied, then this BTD of $\mathcal{Y}$ is unique up to 
	scaling and permutation ambiguities. 
\end{theorem}

\emph{Remark 1:} Note that $k^{'}_A$ is the generalized k-rank 
	concept of the concatenated matrix $\boldsymbol{A}$, which is defined as 
	the maximal number $k^{'}_A$ such that any set of $k^{'}_A$ submatrices 
	of $\boldsymbol{A}$ yields a set of linearly independent columns. 
	
\emph{Remark 2:}	Specifically, if there exists an alternative BTD solution $(\bar{\boldsymbol{A}}
	,\bar{\boldsymbol{B}},\bar{\boldsymbol{C}})$ which also decomposes 
	$\mathcal{Y}$ into $Q$ rank-$(L_q,L_q,1)$ terms, then we have 
	$\bar{\boldsymbol{A}} = \boldsymbol{A}\boldsymbol{\Lambda}_A\boldsymbol{\Pi}$, 
	$\bar{\boldsymbol{B}} = \boldsymbol{B}\boldsymbol{\Lambda}_B\boldsymbol{\Pi}$, 
	and $\bar{\boldsymbol{C}} = \boldsymbol{C}\boldsymbol{\Lambda}_C\boldsymbol{\Pi}_C$, 
	where $\boldsymbol{\Pi}$ is a $Q$-block permutation matrix; $\boldsymbol{\Pi}_C$  
	is a permutation matrix whose permutation pattern is the same as that of  
	$\boldsymbol{\Pi}$; $\boldsymbol{\Lambda}_A$ and $\boldsymbol{\Lambda}_B$ 
	are nonsingular $Q$-block diagonal matrices; $\boldsymbol{\Lambda}_C$ is 
	a nonsingular diagonal matrix. Note that the block structure of $\boldsymbol{\Pi}$, 
	$\boldsymbol{\Lambda}_A$ and $\boldsymbol{\Lambda}_B$ is compatible with 
	that of $\boldsymbol{A}$ and $\boldsymbol{B}$. Let $\boldsymbol{\Lambda}_{A,q}$  
	and $\boldsymbol{\Lambda}_{B,q}$ denote the $q$-th diagonal block of 
	$\boldsymbol{\Lambda}_{A}$ and $\boldsymbol{\Lambda}_{B}$, respectively, 
	and $\lambda_{q}$ denote the $q$-th diagonal element of $\boldsymbol{\Lambda}_{C}$. 
	Then, we have $\lambda_{q}\boldsymbol{\Lambda}_{A,q}\boldsymbol{\Lambda}_{B,q}^T
	=\boldsymbol{I}_{L_q},\forall q$.

According to Theorem \ref{thero_gkkc}, the essential uniqueness of the BTD formulation 
in (\ref{y_tensor3}) is guaranteed if the following two sufficient conditions are met
\begin{align}
	PM \geq \sum\nolimits_{k=1}^{K}L_k^2,
	\label{con1}\\
	k^{'}_G + k^{'}_A + k_{S_o} \geq 2 K + 2.
	\label{con2}
\end{align}
The first dimensionality constraint (\ref{con1}) is generally non-restrictive in 
mmWave/THz ELAA systems, since the number of paths is usually small relative to 
the other two dimensions. Following an analysis similar to that in the previous section, 
we can arrive at $k^{'}_G = K$ and $k^{'}_A = K$ with probability one. {Substituting 
this into (\ref{con2}), the uniqueness condition reduces to $k_{S_o} \geq 2$, 
which can be satisfied by designing pairwise independent training sequences
associated with different users (with the minimum possible value being $T \geq 2$).}

\subsection{Joint Path-Number Estimation and BTD Initialization}
In the considered BTD model, the number of users is usually known a \emph{priori}, 
whereas the number of propagation paths associated with each user is generally unknown. 
Hence, a path number estimation step is required before performing the BTD.
For the $k$-user, let the corresponding BTD block matrix be denoted by $\boldsymbol{X}_k = 
\boldsymbol{G}_k^T\boldsymbol{A}_k$. Since this block is generated by a limited number 
of propagation paths, its effective rank provides a natural estimate for the number of 
paths associated with the $k$-th user.
Meanwhile, the pilot matrix $\boldsymbol{S}_o$ is known at the BS, which enables the 
received tensor to be reformulated as a pilot-aided matrix regression model:
\begin{align}
	\boldsymbol{Y}_{(3)} = \boldsymbol{X}_{(3)} \boldsymbol{S}_o^{T} + 
	\boldsymbol{N}_{(3)}, \label{btd_reg}
\end{align}
where the $k$-th column of $\boldsymbol{X}_{(3)}$ satisfies $\boldsymbol{X}_{(3)}
(:,k) = \text{vec}(\boldsymbol{X}_k)$. 

Based on the regression model in \eqref{btd_reg} and the low-rank property of each 
user-specific block, the joint path-number estimation and BTD initialization can be 
addressed through the following block-wise low-rank recovery problem:
\begin{align}
	\min_{\{\boldsymbol{X}_k\}} \quad &\frac{1}{2}\|\boldsymbol{Y}_{(3)} - \boldsymbol{X}_{(3)}\boldsymbol{S}_o^T\|_F^2 + \lambda \sum\nolimits_{k=1}^{K}
     \|\boldsymbol{X}_k\|_{*}
    \nonumber\\
    \text{s.t.} \quad& \boldsymbol{X}_{(3)}(:,k) = \text{vec}(\boldsymbol{X}_k),
    \label{p_bt1}
\end{align}
where $\|\cdot\|_{*}$ denotes the nuclear norm and $\lambda \ge 0$ controls the strength 
of the low-rank regularization. The data-fitting term enforces the consistency with 
the observed signal and the known pilot sequences, while the nuclear-norm penalty promotes 
the low-rank structure of each user-specific block. 

To solve \eqref{p_bt1}, we introduce auxiliary variables $\{\boldsymbol{Z}_k\}_{k=1}^{K}$, 
and then the problem can be equivalently written as 
\begin{align}
	\min_{\{\boldsymbol{X}_k\},\{\boldsymbol{Z}_k\}} \quad &\frac{1}{2}\|\boldsymbol{Y}_{(3)} - \boldsymbol{X}_{(3)}\boldsymbol{S}_o^T\|_F^2 + \lambda \sum\nolimits_{k=1}^{K}
	\|\boldsymbol{Z}_k\|_{*}
	\nonumber\\
	\text{s.t.} \quad& \boldsymbol{X}_{(3)}(:,k) = \text{vec}(\boldsymbol{Z}_k),
	\label{p_bt2}
\end{align}
The above nuclear-norm-regularized least-squares (LS) problem can be efficiently solved by 
an ADMM framework \cite{boyd2011distributed}, where $\boldsymbol{X}_{(3)}$ is updated by solving the
LS subproblem, and $\boldsymbol{Z}_k$ is updated by solving the nuclear-norm proximal 
subproblem through the singular-value-thresholding (SVT) method \cite{cai2010singular}.

After convergence, the number of paths associated with the $k$-th user can be estimated from the 
effective rank of the recovered block $\hat{\boldsymbol{X}}_k$, i.e. $\hat{L}_k = 
\text{rank}(\hat{\boldsymbol{X}}_k)$. Let $\hat{\boldsymbol{X}}_k = 
\boldsymbol{U}_k\boldsymbol{\Lambda}_k\boldsymbol{V}_k^H$ denote the SVD of 
$\hat{\boldsymbol{X}}_k$, the BTD factor matrices are initialized as
\begin{align}
	&\hat{\boldsymbol{G}}_k^{(0)} = \boldsymbol{U}_{k}(:,1:\hat{L}_k)
	\boldsymbol{\Lambda}_k^{1/2}(1:\hat{L}_k,1:\hat{L}_k),\nonumber\\
	&\hat{\boldsymbol{A}}_k^{(0)} = \boldsymbol{V}_{k}^{*}(:,1:\hat{L}_k)
	\boldsymbol{\Lambda}_k^{1/2}(1:\hat{L}_k,1:\hat{L}_k),\nonumber\\
	&\hat{\boldsymbol{s}}_k^{(0)} = \boldsymbol{s}_k,
	\label{btd_ini}
\end{align}
which provides a reliable starting point for the subsequent BTD refinement, 
and thereby significantly accelerates the convergence of the NLS algorithm.

\subsection{BT Decomposition}
The BTD can be accomplished by solving the following optimization problem 
\begin{align}
	\min_{\{\boldsymbol{G}_k,\boldsymbol{A}_k,\boldsymbol{s}_k\}}
	\quad \frac{1}{2} \|\mathcal{Y} - \sum\nolimits_{k=1}^{K}(\hat{\boldsymbol{G}}_k  
	\hat{\boldsymbol{A}}_k^T) \circ  \hat{\boldsymbol{s}}_{k}\|_F^2.
	\label{p_btd1}
\end{align}
The above problem can be efficiently solved by 
the non-linear least squares (NLS) algorithm implemented in Tensorlab 
\cite{vervliet2016tensorlab}, which jointly updates all factor 
matrices in an iterative manner by exploiting the Jacobian structure of the 
tensor model, typically using a Gauss-Newton (GN) \cite{paatero1999gn} or 
Levenberg-Marquardt (LM) strategy \cite{nion2006levenberg}. {Compared with the ALS algorithm, 
which updates one factor matrix at one time while fixing the others, the NLS algorithm 
takes into account the strong coupling among the factor matrices in the BTD model and is less 
prone to suboptimal convergence \cite{de2008decompositions3,de2008decompositions}.} To be concrete, 
the problem (\ref{p_btd1}) can be re-written as 
\begin{align}
	\min_{\boldsymbol{m}} \quad \frac{1}{2}\boldsymbol{\mu}^H(\boldsymbol{m})
	\boldsymbol{\mu}(\boldsymbol{m}),
\end{align}
where $\boldsymbol{m}$ denotes the variable vector that contains the elements of 
all $K$ block-terms $\{\hat{\boldsymbol{G}}_k,\hat{\boldsymbol{A}}_k,
\hat{\boldsymbol{s}}_k\}$ and $\boldsymbol{\mu}(\boldsymbol{m})$ denotes the 
vector of residuals, which is given by 
\begin{align}
	\boldsymbol{\mu}(\boldsymbol{m})\triangleq
	\text{vec}\left(\mathcal{Y} - \sum\nolimits_{k=1}^{K}(\hat{\boldsymbol{G}}_k  
	\hat{\boldsymbol{A}}_k^T) \circ  \hat{\boldsymbol{s}}_{k}\right) \in \mathbb{C}^{MPT}.
\end{align}
Then, the Jacobian matrix at the point $\boldsymbol{m}^{(t)}$ can be computed by 
$\boldsymbol{J}^{(t)} = \frac{\partial \boldsymbol{\mu}(\boldsymbol{m}^{(t)})}
{\partial \boldsymbol{m}^{(t)}}$,
and the parameter vector can be updated iteratively using the GN or LM step as 
follows 
\begin{align}
	\boldsymbol{m}^{(t+1)} =  \boldsymbol{m}^{(t)} + 
	\left(
 \big(\boldsymbol{J}^{(t)}\big)^H\boldsymbol{J}^{(t)}+
 \lambda^{(t+1)}\boldsymbol{I}\right)^{-1}\big(\boldsymbol{J}^{(t)}\big)^H\boldsymbol{m}^{(t)},
	\label{update_nls}
\end{align}
where $\lambda^{(t+1)}$ denotes the {sampling} parameter at the $(t+1)$-th iteration. Eq.
(\ref{update_nls}) behaves as a GN update for a small value of $\lambda^{(t+1)}$, 
and approaches a LM update when $\lambda^{(t+1)}$ becomes large. 

According to Theorem \ref{thero_gkkc}, the BTD is unique up to scaling and 
permutation ambiguities under some mild conditions, i.e., 
\begin{align}
	\hat{\boldsymbol{G}} &= \boldsymbol{G} \boldsymbol{\Lambda}_{1}\boldsymbol{\Pi} 
	+ \boldsymbol{E}_1,
	\nonumber\\
	\hat{\boldsymbol{A}}  &= \boldsymbol{A} \boldsymbol{\Lambda}_{2}\boldsymbol{\Pi} 
	+ \boldsymbol{E}_2,
	\nonumber\\
	\hat{\boldsymbol{S}}_o &= \boldsymbol{S}_o \boldsymbol{\Lambda}_{3}\boldsymbol{\Pi}_{s} 
	+ \boldsymbol{E}_3,
	\label{gkkc1}
\end{align}
where $\boldsymbol{G} = \left[\boldsymbol{G}_1 \ \cdots \ \boldsymbol{G}_K\right]$, 
$\boldsymbol{A} = \left[\boldsymbol{A}_1 \ \cdots \ \boldsymbol{A}_K\right]$, 
$\boldsymbol{\Lambda}_{1} = \text{blkdiag}(\boldsymbol{\Lambda}_{1,1},\cdots,
\boldsymbol{\Lambda}_{1,K})$ and $\boldsymbol{\Lambda}_{2}=\text{blkdiag}
(\boldsymbol{\Lambda}_{2,1},\cdots,\boldsymbol{\Lambda}_{2,K})$ are non-singular 
$K$-block diagonal matrices and $\boldsymbol{\Lambda}_3 =\text{diag}(\lambda_{3,1},
\cdots,\lambda_{3,K})$ is a non-singular diagonal 
matrix, which satisfy $\lambda_{3,k}\boldsymbol{\Lambda}_{1,k}\boldsymbol{\Lambda}
_{2,k}^T = \boldsymbol{I}_{L_k},\forall k = 1,\cdots,K$; $\boldsymbol{\Pi}$ is a $K$-block 
permutation matrix, $\boldsymbol{\Pi}_s$ is a permutation matrix, and both share 
the same permutation pattern; $\boldsymbol{E}_1$, $\boldsymbol{E}_2$ and 
$\boldsymbol{E}_3$ denote the estimation errors of the BTD.

\subsection{Channel Estimation}
After obtaining the estimated factor matrices $(\hat{\boldsymbol{G}},\hat{\boldsymbol{A}},\hat{\boldsymbol{S}})$ via 
the NLS algorithm, we first 
estimate the permutation matrix $\boldsymbol{\Pi}_s$ to establish the 
correspondence between the recovered BTD components and the users. 
By exploiting the knowledge of the pilot sequences, this association 
can be efficiently determined using a correlation-based method, similar 
to that employed in the LoS scenarios. Specifically, 
$\boldsymbol{\Pi}_s$ is a $K \times K$ permutation matrix, and 
$\boldsymbol{\Pi}$ is a block-permutation matrix consisting of $K \times K$
blocks. If the $k$-th column of $\boldsymbol{S}_o$ is highly correlated 
with the $k'$-th column of $\hat{\boldsymbol{S}}_o$, then the $(k',k)$-th element 
of $\boldsymbol{\Pi}_s$ is set to $1$, and the corresponding $(k',k)$-th block 
of $\boldsymbol{\Pi}$ is set to an identity matrix $\boldsymbol{I}_{{\hat{L}_k}}$. 
Otherwise, the corresponding entry of $\boldsymbol{\Pi}_s$ is $0$, and the corresponding 
block of $\boldsymbol{\Pi}$ is set to a all-zeros matrix. 
For instance, consider a 
$3 \times 3$ permutation matrix 
\begin{align}
	\boldsymbol{\Pi}_s = \begin{bmatrix}
		0 &1 &0\\ 
		1 &0 &0 \\ 
		0 &0 &1
	\end{bmatrix}.
\end{align}
Let $\boldsymbol{\Pi} \in \mathbb{R}^{8\times 10}$ be a $3\times 3$ block 
permutation matrix that follows the same permutation pattern, where the 
block sizes are $2,3$ and $4$, respectively. Then, $\boldsymbol{\Pi}$ can be written as
\begin{align}
	\boldsymbol{\Pi} = \begin{bmatrix}
		\boldsymbol{0}_{2\times3} &\boldsymbol{I}_3 & \boldsymbol{0}_{2\times4} \\ 
		\boldsymbol{I}_2 &\boldsymbol{0}_{3\times2} &\boldsymbol{0}_{3\times4} \\ 
		\boldsymbol{0}_{4\times3} &\boldsymbol{0}_{4\times2} &\boldsymbol{I}_4
	\end{bmatrix}.
\end{align}

After that, the permutation ambiguity for the estimated 
factor matrices can be removed. Thus, we have 
\begin{align}
	&\hat{\boldsymbol{G}} = \hat{\boldsymbol{G}}\boldsymbol{\Pi}^T
	= \boldsymbol{G}\boldsymbol{\Lambda}_1 + \boldsymbol{E}_1,
	\nonumber\\
	&\hat{\boldsymbol{A}} =\hat{\boldsymbol{A}}\boldsymbol{\Pi}^T
	= \boldsymbol{A}\boldsymbol{\Lambda}_2 + 
	\boldsymbol{E}_2,
	\nonumber\\
	&\hat{\boldsymbol{S}}_o =\hat{\boldsymbol{S}}_o\boldsymbol{\Pi}_s^T
	= \boldsymbol{S}_o\boldsymbol{\Lambda}_3 + 
	\boldsymbol{E}_3.
	\label{est2_factor}
\end{align}
Similar to (\ref{est_t_s}), the delay parameter $\tau_{l}$ 
associated with the $l$-th path component can be estimated via 
\begin{align}
	\hat{\tau}_l = \arg \max_{\tau}  \frac{\big|\boldsymbol{g}^H
		(\tau)\hat{\boldsymbol{g}}_l\big|}
	{\|\boldsymbol{g}(\tau)\|_2\|\hat{\boldsymbol{g}}_l\|_2}.
	\label{est_t_m}
\end{align}

{For the multi-path scenario, the distance parameter cannot 
be directly determined from the estimated delays $\{\hat{\tau}_l\}$. 
Let $\hat{\boldsymbol{a}}_l$ denote the $l$-th column of $\hat{\boldsymbol{A}}$, 
which is a scaled version of $\tilde{\boldsymbol{b}}(\theta_l,r_l)=\boldsymbol{W}^H\boldsymbol{b}(\theta_l,r_l)$. 
The angle-distance pair can therefore be jointly estimated via
\begin{align}
	\hat{\theta}_l,\hat{r}_l = \arg \max_{\theta,r}  \frac{\big|
		\boldsymbol{b}^H(\theta,r)\boldsymbol{W}
		\hat{\boldsymbol{a}}_l\big|}
	{\|\boldsymbol{W}^H\boldsymbol{b}(\theta,r)\|_2
	\|\hat{\boldsymbol{a}}_l\|_2}.
	\label{est_thetar}
\end{align}
The optimization in \eqref{est_thetar} is implemented using the same 
multi-level coarse-fine search framework described previously.}
As analyzed in \cite{zhou2017low}, the correlation-based scheme in 
(\ref{est_t_m}) and (\ref{est_thetar}) is a maximum likelihood (ML) 
estimator, provided that the entries of the estimation error matrices
$\boldsymbol{E}_1$ and $\boldsymbol{E}_2$ are i.i.d. complex Gaussian.

Next, we try to recover the path loss gains $\{\hat{\alpha}_l\}$. After 
obtaining $\{\hat{\tau}_l,\hat{\theta}_l,\hat{r}_l\}$, we can reconstruct 
\begin{align}
	\tilde{\boldsymbol{G}} &= [\boldsymbol{g}(\hat{\tau}_1) \ \cdots
	\ \boldsymbol{g}(\hat{\tau}_{L})],
	\nonumber\\
	\tilde{\boldsymbol{B}} &= \boldsymbol{W}^H [\boldsymbol{b}(
	\hat{\theta}_1,\hat{r}_1) \ \cdots \ \boldsymbol{b}(\hat{\theta}_{L},
	\hat{r}_{L})].
\end{align} 
Ignoring the estimation error, the scaling ambiguity matrices can be 
estimated as $\boldsymbol{\Lambda}_1 = \tilde{\boldsymbol{G}}^{\dagger}
\hat{\boldsymbol{G}}$ and $\boldsymbol{\Lambda}_3 = \boldsymbol{S}_o
^{\dagger}\hat{\boldsymbol{S}_o}$. Moreover, since $\lambda_{3,k}\boldsymbol{\Lambda}_{1,k}\boldsymbol{\Lambda}_{2,k}^T 
= \boldsymbol{I}_{{\hat{L}_k}},\forall k =1,\cdots,K$, we have 
\begin{align}
	\boldsymbol{\Lambda}_2^{\dagger} = \text{blkdiag}(
	\lambda_{3,1}\boldsymbol{\Lambda}_{1,1}^T,\cdots,
		\lambda_{3,K}\boldsymbol{\Lambda}_{1,K}^T).
\end{align}
Ideally we should have $\hat{\boldsymbol{A}} = \tilde{\boldsymbol{B}}
\boldsymbol{D}_{\alpha} \boldsymbol{\Lambda}_2$, where $\boldsymbol{D}
_{\alpha} \triangleq \text{diag}(\alpha_1,\cdots,\alpha_{L})$. 
Hence, $\boldsymbol{D}_{a}$ can be estimated as 
\begin{align}
	\hat{\boldsymbol{D}}_{\alpha} = \text{diag}
	(\tilde{\boldsymbol{B}}^{\dagger}\hat{\boldsymbol{A}}
	\boldsymbol{\Lambda}_2^{\dagger}).
	\label{est_alpha}
\end{align}
Finally, the channels $\{\boldsymbol{h}_{p,k}\}$ can be reconstructed 
by the estimated parameters $\{\hat{\tau}_l,\hat{\theta}_l,\hat{r}_l,
\hat{\alpha}_l\}$ according to (\ref{h_channel}). 

{\textit{Remark 3:}
The BTD-based formulation can also be extended to a more general mixed LoS/NLoS scenario, 
where the LoS paths are unobstructed for some users but blocked for others. Specifically, 
a user with only one dominant LoS path can be regarded as a special case corresponding to 
a rank-one block, while a user with multiple reflected or scattered NLoS paths corresponds 
to a higher-rank block.}

{ For clarity, the overall processing flow of the proposed BTD-based framework is 
	summarized as Algorithm \ref{alg_btd}.}
{\begin{algorithm}[H]
\caption{BTD-based Channel Estimation Framework}
		\begin{algorithmic}[1]			
			\REQUIRE{$\mathcal{Y}$, $K$, $\boldsymbol{S}_o$, $\boldsymbol{W}$.}
			\ENSURE{$\{\hat{\boldsymbol{h}}_{p,k}\}_{k=1,p=1}^{K,P}$.}
			\STATE Estimate the number of paths for each user and initialize 
			the factor matrices by solving \eqref{p_bt1}.
			\STATE Perform BTD via the NLS algorithm.
			\STATE Determine user-path association using a correlation-based method and 
			remove the permutation ambiguity of the recovered factor matrices.
			\STATE Extract the channel parameters $\{\hat{\tau}_l,\hat{\theta}_l,\hat{r}_l\}$
			through the multi-level coarse-fine search in \eqref{est_t_m}--\eqref{est_thetar}.
			\STATE Estimate the path gains $\{\hat{\alpha}_l\}$ according to \eqref{est_alpha}.
			\STATE Reconstruct the multi-user channels.
		\end{algorithmic}
		\label{alg_btd}
\end{algorithm}}

{
\section{Computational Complexity and Convergence Analysis}

\subsection{Computational Complexity Analysis}
For the CPD-based method, the dominant computational complexity arises from the ALS algorithm, 
in which the three factor matrices are iteratively updated by solving the corresponding LS
subproblems. Ignoring lower-order terms, the overall computational complexity is on the 
order of $\mathcal{O}(PMTKN_{\rm CPD})$, where $N_{\rm CPD}$ denotes the number of ALS 
iterations. 

For the BTD-based method, the dominant computational complexity mainly comes from two tasks, 
namely, ADMM-based path number estimation procedure and the subsequent 
NLS refinement. Specifically, the complexity of the ADMM framework is on the order of 
$\mathcal{O}((PMK^2+PMK\bar{L})N_{\rm ADMM})$, where $N_{\rm ADMM}$ denotes the number of 
ADMM iterations and $\bar{L}$ denotes average number of effective eigenvalues in each block. 
Given the initialized BTD factors, the NLS refinement incurs a complexity on the order of $\mathcal{O}((PMTD^2+D^3)N_{\rm BTD})$,  where $D\triangleq L(P+M)+KT$ denotes the 
dimension of the variable vector, and $N_{\rm BTD}$ denotes the number BTD iterations. 
Note that the above complexity corresponds to a conservative order-level estimate. 
In practice, NLS implementation in Tensorlab exploits structured Jacobians and efficient 
numerical routines, which further reduce the actual computational complexity.

\subsection{Convergence Analysis}
The convergence behavior of both the ALS and NLS algorithms has been extensively 
studied in the tensor decomposition literature \cite{kruskal1977three,conway1996packing,Sorensen2013blind,de2008decompositions,de2008decompositions3}. 
The ALS algorithm updates the 
factor matrices of a CPD model in a mode-wise fashion. In each iteration, one factor matrix is optimized while keeping the others fixed, which reduces to a LS problem. 
Since each update exactly minimizes the corresponding subproblem, the overall CPD objective function is guaranteed to be non-increasing throughout the iterations. 

As for the NLS, it treats all factor matrices as a coupled variable and iteratively minimizes a 
NLS objective. The updates are typically obtained using GN or LM schemes, which generate descent directions for the objective function. 
Consequently, the objective function is also non-increasing under standard NLS update rules. 
Since the objective functions of both algorithms are monotonically non-increasing during the 
iterative process, both algorithms are guaranteed to converge to a stationary point.}

\section{Simulation Results}
In this section, we present simulation results to illustrate the performance 
of the proposed CPD/BTD-based near-field channel estimation and sensing method. For the LoS scenario, since the 
dominant LoS path provides reliable geometric information, the recovered channel 
parameters allow not only accurate channel estimation but also precise user localization. 
Therefore, both channel estimation performance and localization accuracy are evaluated. 
For the NLoS scenario, we only focus on evaluating the  
channel estimation performance of the proposed method. 

In our simulations, the signal-to-noise ratio (SNR) is defined as 
\begin{align}
	\text{SNR} \triangleq
	\|\mathcal{Y}-\mathcal{N}\|_F^2/\|\mathcal{N}\|_F^2.
\end{align}
We also derive the Cram\'{e}r–Rao bounds (CRBs) for both the channel parameters and the 
user positions, which serve as a theoretical lower bound for evaluating the estimation accuracy of 
channel parameters and user locations. The normalized mean square error (NMSE) is used to 
quantify the channel estimation performance, which is defined as
\begin{align}
	\text{NMSE} \triangleq \frac{
		\sum_{p=1}^{P}\sum_{k=1}^{K}\|\boldsymbol{h}_{p,k}-\hat{\boldsymbol{h}}_{p,k}\|_F^2}
	{\sum_{p=1}^{P}\sum_{k=1}^{K}\|\boldsymbol{h}_{p,k}\|_F^2}.
\end{align}

The pilot symbol matrix $\boldsymbol{S}_{o}$ is chosen from the codebook of Grassmannian 
beamforming \cite{love2003grassmannian} for $T = 2$. While for $3\leq T < 8$, 
$\boldsymbol{S}_{o}$ can be calculated by the algorithm proposed in 
\cite{medra2015flexible}. When $T = 8$, $\boldsymbol{S}_{o}$ is simply chosen as a 
DFT matrix.

\subsection{MMV-Based Channel Estimation Benchmarks}\label{sec:benchmark}
Owing to the sparse scattering nature of mmWave/THz channel, the multi-user 
channel estimation considered in this paper can also be reformulated as a 
multiple-measurement-vector (MMV) compressed sensing problem. Taking the 
mode-1 unfolding of $\mathcal{Y}$ in (\ref{CPD}), we have
\begin{align}
	\boldsymbol{Y}_{(1)}^T &= (\boldsymbol{S} \odot \boldsymbol{A})\boldsymbol{G}^T
	+ \boldsymbol{N}_{(1)}^T
	\nonumber\\
	&= (\boldsymbol{S} \otimes \boldsymbol{W}^H\bar{\boldsymbol{B}})^T(\boldsymbol{O}
	\otimes \boldsymbol{X}_{B})
	\boldsymbol{D}_{\alpha}\boldsymbol{G}^T
	+ \boldsymbol{N}_{(1)}^T,
	\label{y1_tensor}
\end{align}
where $\boldsymbol{Y}_{(1)}$ and $\boldsymbol{N}_{(1)}$ are the mode-1 unfolding 
of $\mathcal{Y}$ and $\mathcal{N}$, respectively; $\bar{\boldsymbol{B}}\in
\mathbb{C}^{N\times Q}$ is the near-field polar-domain codebook \cite{cui2022channel}; 
$\boldsymbol{X}_B\in\mathbb{C}^{Q\times P}$ is a sparse matrix whose different columns 
exhibit the same sparsity pattern. Define 
\begin{align}
	\boldsymbol{\Phi} &\triangleq \boldsymbol{S}
	\otimes \boldsymbol{W}^H \bar{\boldsymbol{B}},
	\nonumber\\
	\boldsymbol{X} &\triangleq 
	(\boldsymbol{O}\otimes \boldsymbol{X}_{B})\boldsymbol{D}_{\alpha}\boldsymbol{G}^T.
\end{align}
It is obvious that $\boldsymbol{X}$ is obtained by augmenting $\boldsymbol{X}_B$ 
with zero rows and thus its columns also exhibit the common sparsity pattern. 
Consequently, the mode-1 unfolding of $\mathcal{Y}$ can be re-written as 
\begin{align}
	\boldsymbol{Y}_{(1)}^T = \boldsymbol{\Phi} \boldsymbol{X} + \boldsymbol{N}_{(1)}^T.
\end{align}
The above problem can be solved by the simultaneous orthogonal matching pursuit (SOMP)
\cite{cui2022channel} and the simultaneous iterative gridless weighted (SIGW) \cite{cui2022channel} 
algorithms, which are adopted as the benchmarks in our simulations. 

\begin{table}[t]
	\centering
	\caption{Main simulation parameters}
	\label{tab:sim_parameters}
	\small
	\setlength{\tabcolsep}{8.2pt}
	\renewcommand{\arraystretch}{1}
	\begin{tabular}{c|c|c}
		\hline
		\textbf{Parameter} & \textbf{LoS} & \textbf{NLoS} \\
		\hline
		Carrier freq. $f_c$ & $100$ GHz & $30$ GHz \\
		Bandwidth $B$ & $0.1$ GHz & $0.1$ GHz \\
		Subcarriers $P$ & $64$ & $64$ \\
		BS antennas $N$ & $256$ & $128$ \\
		RF chains $M$ & $32^{\ast}$ & $64^{\ast}$ \\
		Users $K$ & $8$ & $8$ \\
		Pilot length $T$ & $4^{\ast}$ & $4^{\ast}$ \\
		Distance range & $[20,80]$ m & $[20,80]$ m \\
		Angle range & $[-60^\circ,60^\circ]$ & $[-60^\circ,60^\circ]$ \\
		\hline
	\end{tabular}\\
	\vspace{0.5mm}
	\footnotesize
	$^{\ast}$ Default value unless otherwise specified.
\end{table}

\subsection{LoS Scenarios (THz Bands)}
We first consider an ELAA THz system where the propagation is dominated by a 
strong LoS component due to the limited diffraction and severe path loss at THz 
frequencies. {Simulation parameters are summarized in Table~\ref{tab:sim_parameters}. 
The users are randomly distributed according to the distance and angle settings in 
Table~\ref{tab:sim_parameters}, and the corresponding LoS delay is computed as $\tau_k=r_k/c$.}
The Rayleigh distance of the considered system is $d_R = 97.5$ meters. 
The gain of the LoS path is generated as \cite{ning2023beamforming}:
\begin{align}
	\alpha_{\text{LoS}} = \frac{c}{4\pi fd}\exp \left(
	-\frac{1}{2}K(f)d\right)\exp(-j2\pi f\tau)
\end{align}
where $f$ and $d$ denote the frequency and the propagation distance, respectively. 
The time delay of the LoS path is given by $\tau = (d/c)$. The molecular 
absorption coefficient, $K(f)$, characterizes the frequency-selective attenuation 
caused by molecular energy conversion in the propagation medium \cite{ning2023beamforming}
and is set to $0.01$ in our experiment. 

Note that after the factor matrices are estimated, there are two different strategies 
in estimating the angle and distance parameters in the LoS scenarios. {The first 
approach jointly estimates the distance and angle parameters from the estimated factor matrix 
$\hat{\boldsymbol{A}}$ via \eqref{est_thetar1}. This approach involves a multi-level 
coarse-fine search over a two-dimensional near-field codebook. 
The second approach, instead, recovers the distance from the estimated delay using 
the geometric relationship $\hat{r}_k=\hat{\tau}_{k}c$. With the distance treated as known, 
the angle $\theta_k$ can be then estimated via a one-dimensional multi-level 
search in (\ref{est_theta}).  
The latter approach generally yields a more accurate estimate of the distance. 
This is because the factor matrix $\boldsymbol{G}$ has a Vandermonde structure which provides 
a better resolvability for different distances, while the near-field steering vectors associated
with different distances may exhibit high correlation, which limits its distance resolution. 
In our simulations, the first approach is referred to as
``CPD-Joint'', while the latter approach is specifically termed as ``CPD-Delay-Aided''.
The corresponding CRB derivations for these two estimation strategies are provided in 
Appendix~A and Appendix~B, respectively.}


\subsubsection{Evaluation of Channel Parameter Estimation Performance}
\begin{figure*}[h]
	\centering
	\begin{subfigure}[b]{0.3\textwidth}
		\includegraphics[width=1\linewidth]{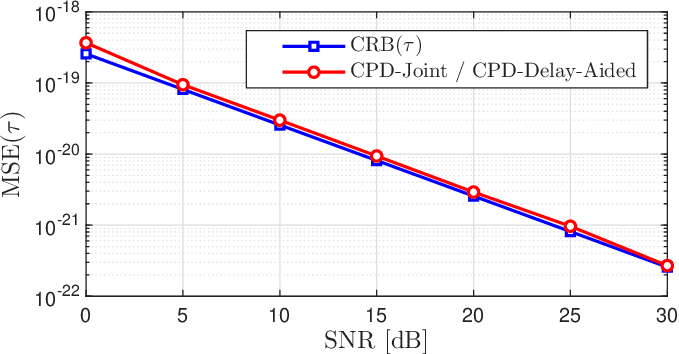}
	\end{subfigure}
	\hfill
	\begin{subfigure}[b]{0.3\textwidth}
		\includegraphics[width=1\linewidth]{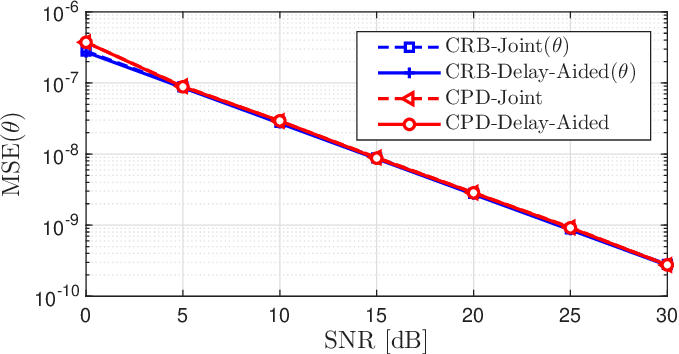}
	\end{subfigure}
	\hfill
	\begin{subfigure}[b]{0.3\textwidth}
		\includegraphics[width=1\linewidth]{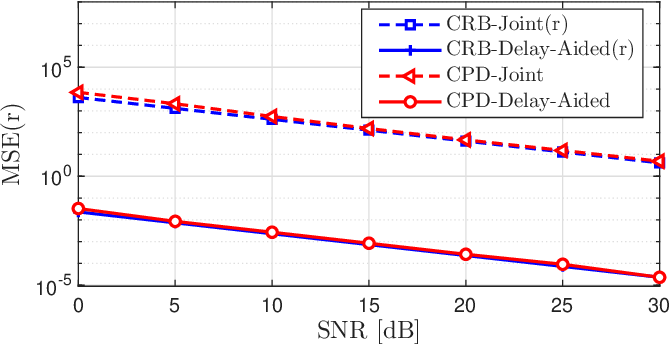}
	\end{subfigure}
	\caption{ MSEs/CRBs for channel parameter estimation versus SNR [dB], where 
	$T = 4$ and $M = 32$.}
	\label{mse_crb_snr256}
\end{figure*}

In Fig. \ref{mse_crb_snr256}, we plot the mean square errors (MSEs) of the estimated channel 
parameters $\{\boldsymbol{\tau},\boldsymbol{\theta},\boldsymbol{r}\}$ as a function of SNR, 
where we set $T = 4$, $P = 64$ and $M = 32$. {It can be observed that the MSEs attained by both 
``CPD-Joint'' and  ``CPD-Delay-Aided'' method gradually approach their corresponding CRBs as 
the SNR increases.} On the other hand, as explained earlier, the {``CPD-Delay-Aided''} method 
can achieve a much more accurate estimate of the distance parameter by utilizing the delay information.

\subsubsection{Evaluation of User Localization Performance}
\begin{figure*}[h]
	\centering
	\begin{subfigure}[b]{0.24\textwidth}
		\centering
		\includegraphics[width=\linewidth]{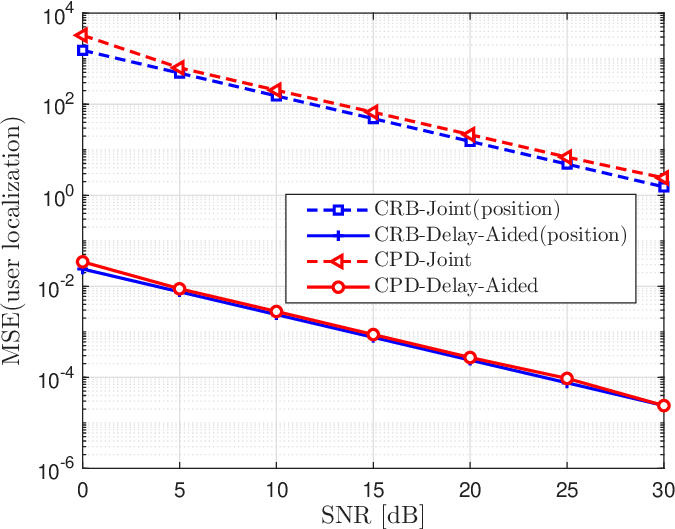}
	\end{subfigure}
	\hfill
	\begin{subfigure}[b]{0.24\textwidth}
		\centering
		\includegraphics[width=\linewidth]{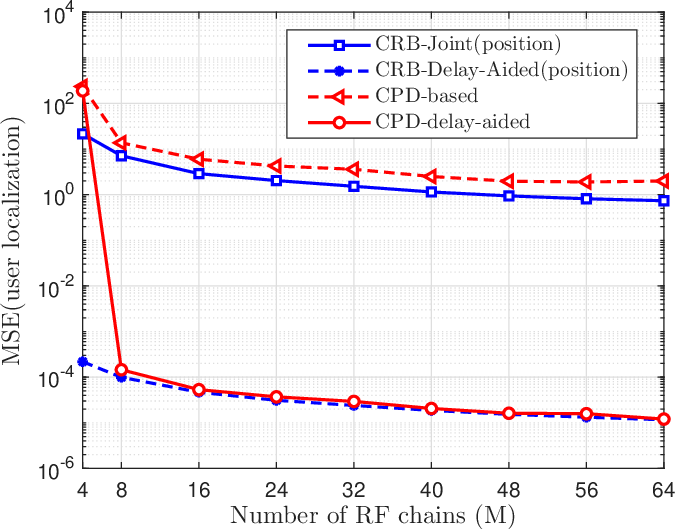}
	\end{subfigure}
	\hfill
	\begin{subfigure}[b]{0.24\textwidth}
		\centering
		\includegraphics[width=\linewidth]{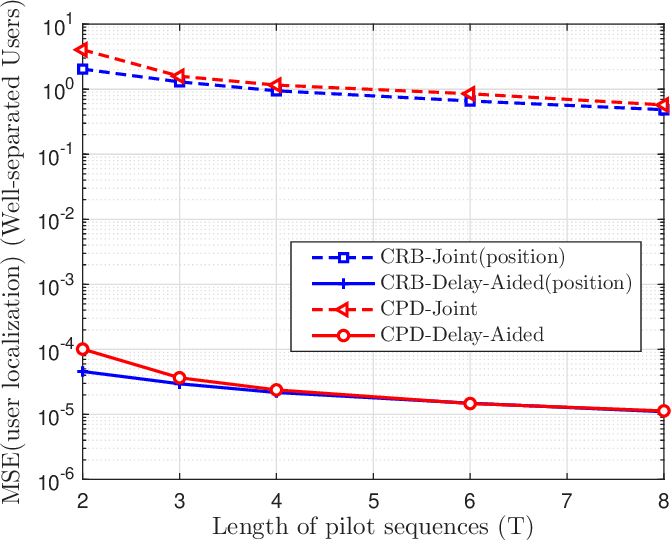}
	\end{subfigure}
	\hfill
	\begin{subfigure}[b]{0.24\textwidth}
		\centering
		\includegraphics[width=\linewidth]{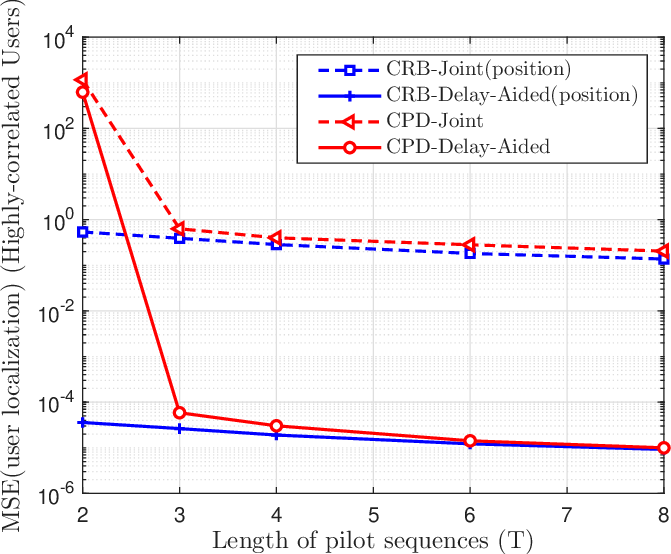}
	\end{subfigure}
	\caption{ MSE/CRB for user localization versus (a) SNR [dB], where $T = 4$ and 
		$M = 32$; (b) the number of RF chains ($M$), where $T=4$ and SNR = $30$ dB; 
		(c) the length of pilot sequences ($T$) for spatially well-separated users, where $M = 32$ and 
		SNR $= 30$ dB;
		(d) the length of pilot sequences ($T$) for spatially weakly separated users, where $M = 32$ and 
		SNR $= 30$ dB.}
	\label{mse_crb_position}
\end{figure*}

{As shown in Fig. \ref{mse_crb_position}, the MSEs of both 
``CPD-Joint'' and  ``CPD-Delay-Aided'' methods are clearly lower-bounded by their corresponding CRBs. }
In Fig. \ref{mse_crb_position}(a), it can be observed that the proposed {``CPD-Delay-Aided''} 
method is capable of achieving a millimeter-level localization accuracy when SNR is above $10$ dB. 
In contrast, {the ``CPD-Joint'' method exhibits substantially higher CRBs and MSEs than the CPD-Delay-Aided'' method.} The primary reason, as discussed before, is that the phase 
of a near-field spherical wavefront varies slowly with distance, making the coherence of near-field 
steering vectors less sensitive to the variation of distance. 
As a result, it becomes highly challenging to obtain a high-precision estimate of the distance 
simply from the factor matrix $\hat{\boldsymbol{A}}$, which in turn affects the user localization 
performance. 

Fig. \ref{mse_crb_position}(b) illustrates the MSEs of estimated user locations versus the number of RF chains. 
We see that the MSEs of {the proposed CPD-based methods}
converge quickly to their CRBs, demonstrating that the proposed method 
can achieve high-accuracy user localization performance even with a 
relatively small number of RF chains. 

{Figs. \ref{mse_crb_position}(c) and (d) depict the MSEs versus the length of pilot 
sequences for spatially well-separated and spatially weakly separated users, respectively. 
To generate these two user-separation scenarios, we first randomly generate a large number of 
user-location realizations and compute the maximum mutual coherence among different
near-field steering vectors. Then, those realizations with maximum mutual coherence greater than $0.7$ 
are selected as the scenarios of spatially weakly separated user, while those realizations with maximum mutual coherence less than $0.4$ are selected as the scenarios of spatially well-separated user.
For users that are well-separated, the proposed CPD-based methods achieve 
superior localization performance even with only $T=2$ pilot symbols. When users become spatially 
highly correlated, the performance degrades due to ill-conditioned factor matrices, in which case  
increasing the pilot length can improve the CPD identifiability and the subsequent estimation accuracy.
Overall, the proposed framework remains effective even when the number of pilot symbols is 
much smaller than the number of users, demonstrating its potential in substantially reducing 
the training overhead.}   

 
\begin{figure*}[h]
	\centering
	\begin{subfigure}[b]{0.24\textwidth}
		\centering
		\includegraphics[width=\linewidth]{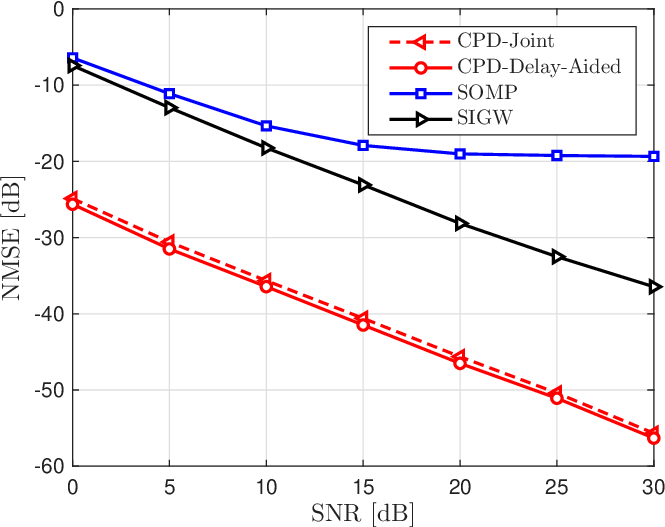}
	\end{subfigure}
	\hfill
	\begin{subfigure}[b]{0.24\textwidth}
		\centering
		\includegraphics[width=\linewidth]{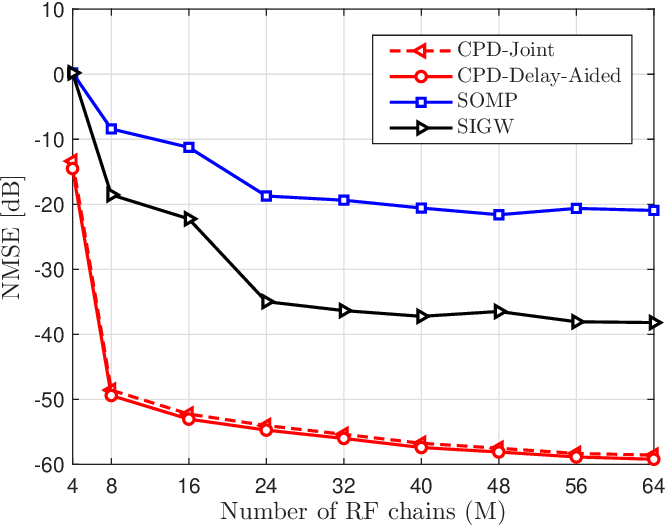}
	\end{subfigure}
	\hfill
	\begin{subfigure}[b]{0.24\textwidth}
		\centering
		\includegraphics[width=\linewidth]{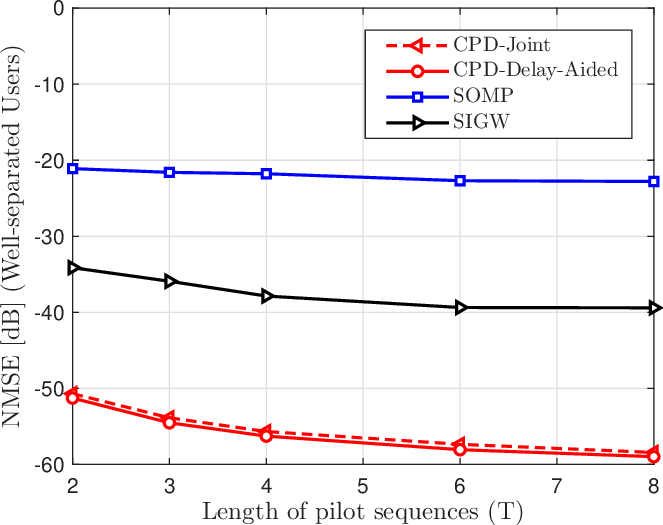}
	\end{subfigure}
	\hfill
	\begin{subfigure}[b]{0.24\textwidth}
		\centering
		\includegraphics[width=\linewidth]{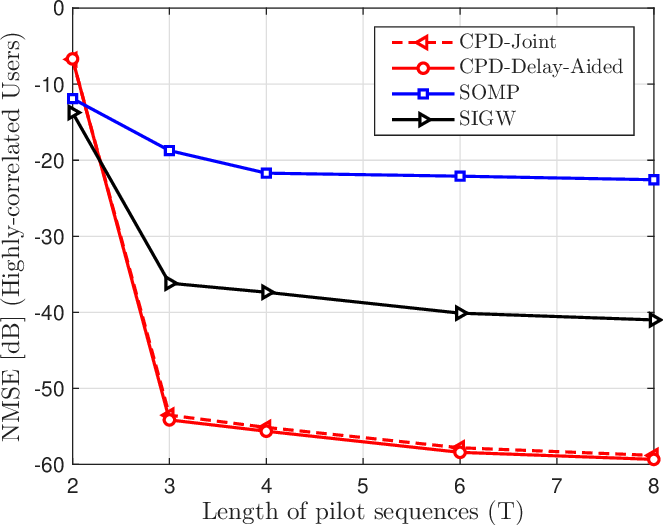}
	\end{subfigure}
	\caption{ NMSE versus (a) SNR [dB], where $T = 4$ and 
		$M = 32$; (b) the number of RF chains ($M$), where $T=4$ and SNR = $30$ dB; 
		(c) the length of pilot sequences ($T$) for spatially well-separated users, where $M = 32$ and 
		SNR $= 30$ dB;
		(d) the length of pilot sequences ($T$) for spatially weakly separated users, where $M = 32$ and 
		SNR $= 30$ dB.}
	\label{nmse_256}
\end{figure*}

\subsubsection{Evaluation of Channel Estimation Performance}
Fig. \ref{nmse_256} plots the NMSE performance of respective algorithms. 
It can be observed that both {``CPD-Joint'' and ``CPD-Delay-Aided''} methods 
present a significant performance improvement over the MMV-based baselines. 
This is attributed to the fact that the proposed methods effectively explore the intrinsic 
multi-dimensional low-rank structure of near-field OFDM channels. The {``CPD-Delay-Aided''}
method slightly outperforms the {``CPD-Joint''} method owing to its more accurate 
estimation of the distance parameters. 

\subsubsection{Evaluation of Computational Complexity}
\begin{table}[t]
	\centering
	\caption{Average run times (s) of respective algorithms 
		under different LoS simulation settings.}
	\label{tab:cpd_runtime_settings}
	\small
	\setlength{\tabcolsep}{5pt}
	\renewcommand{\arraystretch}{1.15}
	\begin{tabular}{c c c c c c c c}
		\hline
		$T$ & $M$ &$P$& SNR & 
		\makecell{CPD-\\Joint} & 
		\makecell{CPD-\\Delay-Aided} & 
		SOMP & SIGW \\
		\hline
		4 & 32 & 32 & 30 dB & 0.4720 & 0.3459 & 0.6910 & 0.8652 \\
		4 & 32 & 64 & 30 dB & 0.5272 & 0.4045 & 1.1169 & 1.2384 \\
		4 & 64 & 64 & 30 dB & 1.0811 & 0.8942 & 1.5344 & 2.3555 \\
		6 & 64 & 64 & 30 dB & 1.1303 & 0.9573 & 1.9005 & 3.3254 \\
		\hline
	\end{tabular}
\end{table}
{Table~\ref{tab:cpd_runtime_settings} compares the average run times of respective algorithms 
under several representative LoS simulation settings. We see that both ``CPD-Joint'' and  ``CPD-Delay-Aided'' methods exhibit lower run time than the MMV-based benchmarks. 
The main reason is that the CPD-based methods only require a limited number of ALS iterations, 
whereas the MMV-based methods have to perform sparse recovery over a large size of near-field polar-domain codebook, leading to a higher computational cost.
In particular, the ``CPD-Delay-Aided'' method further reduces the run time by exploiting 
the delay-distance relationship to avoid the two-dimensional joint angle-distance search required 
in the ``CPD-Joint'' method. As the system dimensions $(P,M,T)$ increase, the run times of the 
MMV-based methods grow rapidly due to the enlarged codebook size. The SIGW algorithm experiences 
an even sharper increase because it additionally relies on iterative gradient-based 
optimization.}

\subsection{NLoS Scenarios (MmWave Bands)}
To demonstrate the effectiveness of our proposed method for more general settings,
we consider NLoS scenarios where the LoS path is blocked between each user and the BS, 
and there exist multiple NLoS paths from the user to the BS. 
{Simulation parameters are summarized in Table~\ref{tab:sim_parameters}.} 
Accordingly, the Rayleigh distance is given by $d_R = 80.6$ m. 
The number of NLoS paths between the BS and each user is randomly 
selected from $\{1,2\}$ and the complex path gain of each NLoS path 
follows $\alpha_{k,l} \sim \mathcal{CN}(0,10^{-0.1(\kappa+\mu)})$, where 
$\kappa = a + 10b\log_{10}(d) + \epsilon$, with $d$ being the distance between the BS 
and the $k$-th user, $\mu = 7$ dB denoting the Rician factor and $\epsilon \in \mathcal{CN}(0,\sigma_{\epsilon}^2)$ \cite{sun2018propagation}. 
Here, we set $a = 61.4$, $b=2$ and $\sigma_{\epsilon} = 5.8$ dB as suggested in 
\cite{sun2018propagation}. The time delay associated with 
the $(k,l)$-th path is given by $\tau_{k,l} = \tilde{\tau}_{k,l} + r_{k,l}/c$, where 
$\tilde{\tau}_{k,l}$ represents the propagation delay from the $k$-th user to the 
scatterer of the $(k,l)$-th path and is randomly chosen from $[0, 2\times10^{-9}]$ 
seconds(s). 
\begin{figure*}[h]
	\centering
	\begin{subfigure}[b]{0.24\textwidth}
		\centering
		\includegraphics[width=\linewidth]{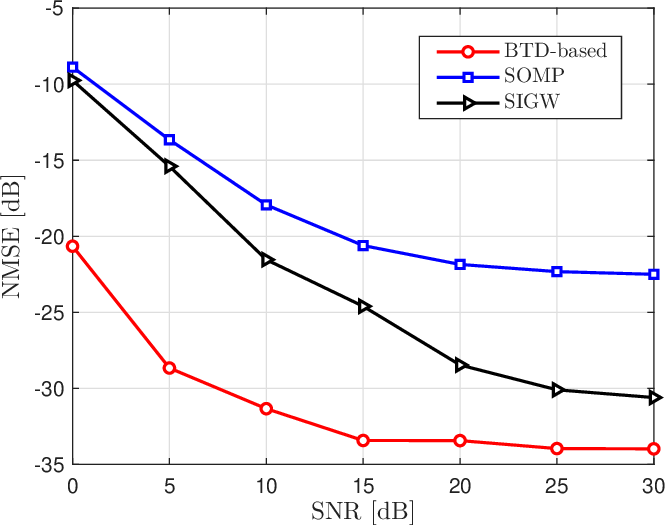}
	\end{subfigure}
	\hfill
	\begin{subfigure}[b]{0.24\textwidth}
		\centering
		\includegraphics[width=\linewidth]{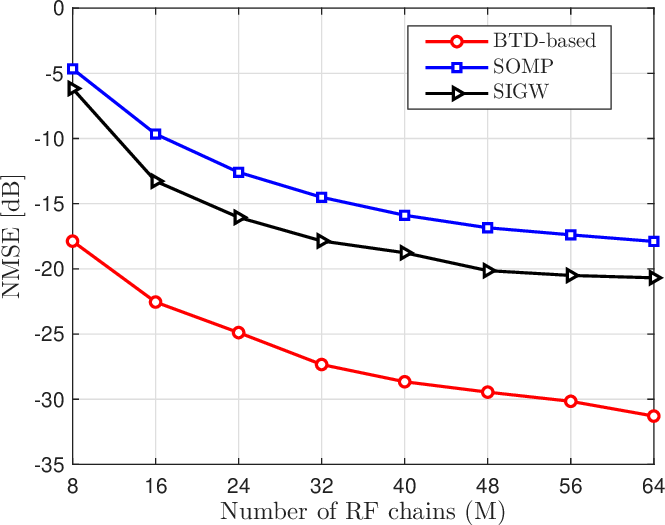}
	\end{subfigure}
	\hfill
	\begin{subfigure}[b]{0.24\textwidth}
		\centering
		\includegraphics[width=\linewidth]{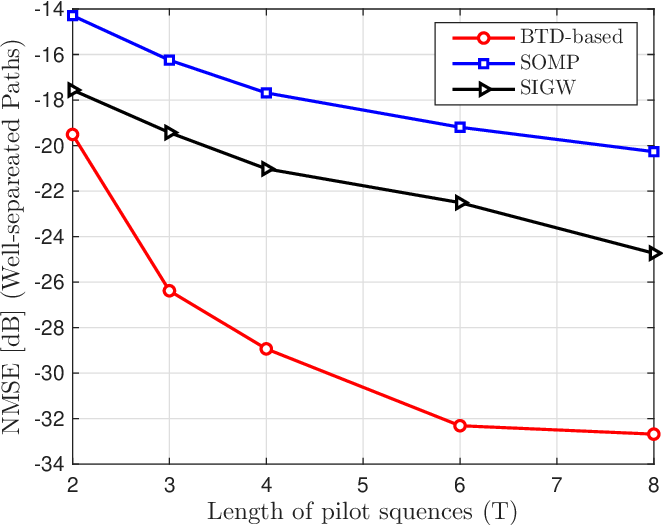}
	\end{subfigure}
	\hfill
	\begin{subfigure}[b]{0.24\textwidth}
		\centering
		\includegraphics[width=\linewidth]{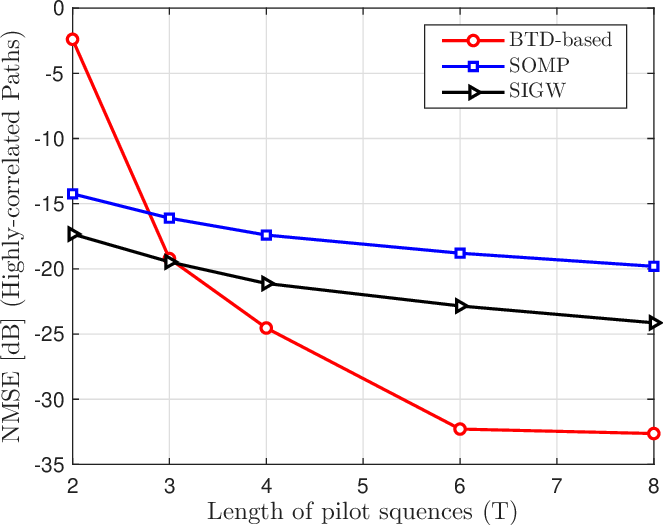}
	\end{subfigure}
	\caption{ NMSE versus (a) SNR [dB], where $T = 4$ and $M = 64$; 
		(b) the number of RF chains ($M$), where $T = 4$ and SNR = $10$ dB; 
		(c) the length of pilot sequences ($T$) for spatially well-separated scenarios, where $M = 64$ and SNR = $10$ dB;
		(d) the length of pilot sequences ($T$) for spatially weakly separated scenarios, 
		where $M = 64$ and SNR = $10$ dB.}
	\label{nmse_128}
\end{figure*}

Fig. \ref{nmse_128} illustrates the NMSE performance of our proposed BTD-based method and the CS-based 
MMV methods, where we set $T = 4$ and $M = 64$. It can be observed that, 
our proposed BTD-based method outperforms the SOMP method by a big margin across different SNR settings. 
Compared with the SIGW method, a significant performance improvement can be observed at 
the low-SNR regime. {In the high-SNR regime, the proposed BTD-based method reaches its error 
floor due to the numerical ill-conditioning of the NLS algorithm, while the SIGW can still benefit 
from the reduced noise level, so the performance gap becomes smaller.}

{As shown in Fig. \ref{nmse_128}(c) and (d), our proposed method achieves a reliable 
channel estimation performance for spatially well-separated scenarios (i.e., the angle/distance parameters associated with different paths are sufficiently separated) with only $T \geq 2$, since the 
Kruskal condition of BTD can be readily satisfied in this case. However, for spatially weakly 
separated scenarios (i.e., the angle/distance parameters associated with different paths are weakly 
separated), the associated factor matrices become ill-conditioned, leading to 
even worse performance than CS-based methods when $T$ is small. This degradation can be 
mitigated by increasing the pilot length, which improves the overall identifiability of the 
BTD model and thereby enhances the subsequent estimation performance.}

\begin{table}[t]
	\centering
	\caption{Average run times (s) of respective algorithms under 
		different NLoS simulation settings.}
	\label{tab:btd_runtime_settings}
	\small
	\setlength{\tabcolsep}{5pt}
	\renewcommand{\arraystretch}{1.15}
	\begin{tabular}{c c c c c c c }
		\hline
		$T$ & $M$ &$P$& SNR & 
		\makecell{BTD} & 
		SOMP & SIGW \\
		\hline
		4 & 32 & 64 & 10 dB & 1.0317 & 0.4150 & 0.6229 \\
		4 & 64 & 32 & 10 dB & 1.3916 & 0.4069 & 1.1853 \\
		4 & 64 & 64 & 10 dB & 1.6694 & 0.5590 & 1.2849 \\
		6 & 64 & 64 & 10 dB & 1.8638 & 0.9942 & 2.5594 \\
		\hline
	\end{tabular}
\end{table}

{Table~\ref{tab:btd_runtime_settings} compares the average run times of respective algorithms 
under several representative NLoS simulation settings. As observed, the proposed BTD-based 
method incurs a higher computational cost than the CS-based benchmarks in most considered settings. 
Interestingly, when the pilot length increases to $T=6$, the BTD-based 
method becomes more efficient than the SIGW method. This improvement can be attributed to the 
faster convergence behavior of the NLS algorithm with a larger pilot length, while the 
computational cost of the SIGW method increases more rapidly as the data dimension grows.}

\section{Conclusion}
In this paper, we developed a tensor decomposition-based method for integrated near-field multi-user 
channel estimation and user localization in ELAA THz systems. 
To improve scalability with an increasing number of users, we consider a non-orthogonal 
pilot transmission scheme where the pilot length is smaller than the number of users. 
By exploring the intrinsic multi-dimensional low-rank structure of near-field channels, the 
received signal is modeled as a third-order tensor that admits a CPD representation in LoS scenarios 
and a BTD representation in the general multi-path environments. These tensor structures enable effective decoupling of 
path components across different users and facilitate accurate channel estimation and user localization.
Our uniqueness analysis further provides a theoretical guarantee for reliable joint estimation   
under the non-orthogonal pilot transmission, {highlighting its potential in training overhead 
reduction.}
Through extensive simulations, we validated the superiority of the proposed methods compared to the 
CS-based channel estimation methods. 


\begin{appendices}
\section{Derivation of CRBs for CPD-Joint Method}
Consider the $P\times M \times T$ observation tensor $\mathcal{Y}$ in (\ref{CPD}) 
\begin{align}
	\mathcal{Y} = \sum\nolimits_{l=1}^{L} \alpha_l \boldsymbol{g}(\tau_l) \circ
	(\boldsymbol{W}^H \boldsymbol{b}(\theta_l,r_l)) \circ \tilde{\boldsymbol{s}}_l
	+ \mathcal{N},
	\label{y_all}
\end{align}
where $\mathcal{N}(m,p,t) \sim \mathcal{CN}(0,\sigma^2)$, $\{\alpha_l,\tau_l,\theta_l,
r_l\}$ are the unknown channel parameters to be estimated. 
Let $\boldsymbol{\xi} \triangleq \left[\boldsymbol{\theta}^T \ 
\boldsymbol{r}^T \ \boldsymbol{\tau}^T \ \boldsymbol{\alpha}_{R}^T \ 
\boldsymbol{\alpha}_{I}^T\right]^T 
\in \mathbb{R}^{5K}$ denote the parameter vector, where 
\begin{align}
	\boldsymbol{\theta} &\triangleq \left[\theta_1 \ \cdots \ \theta_K\right]^T 
	\in \mathbb{R}^{K},
	\nonumber\\
	\boldsymbol{\tau} & \triangleq \left[\tau_1 \ \cdots \ \tau_K\right]^T
	\in \mathbb{R}^{K},
	\nonumber\\
	\boldsymbol{r} &\triangleq \left[r_1 \ \cdots \ r_K\right]^T
	\in \mathbb{R}^{K},
	\nonumber\\
	\boldsymbol{\alpha}_{R}&\triangleq \left[\text{Re}\{\boldsymbol{\alpha}_1\} \ 
	\cdots \ \text{Re}\{\boldsymbol{\alpha}_K\}\right]^T \in \mathbb{R}^{K},
	\nonumber\\
	\boldsymbol{\alpha}_{I}&\triangleq \left[\text{Im}\{\boldsymbol{\alpha}_1\} \ 
	\cdots \ \text{Im}\{\boldsymbol{\alpha}_K\}\right]^T \in \mathbb{R}^{K}.
\end{align} 
 
By stacking the tensor entries in (\ref{y_all}) into a vector form, the received  
signal can be re-written as 
\begin{align}
	\boldsymbol{y}_v =\boldsymbol{\mu}(\boldsymbol{\xi}) + \boldsymbol{n}_v \in \mathbb{C}^{MPT},
\end{align}
where $\boldsymbol{y}_v \triangleq \text{vec}(\mathcal{Y})$, 
$\boldsymbol{n}_v\triangleq \text{vec}(\mathcal{N}) \in \mathbb{C}^{MPT}$ and
\begin{align}
	\boldsymbol{\mu}(\boldsymbol{\xi}) \triangleq \sum\nolimits_{l=1}^{L}
	\alpha_l (\tilde{\boldsymbol{s}}_l \otimes (\boldsymbol{W}^H 
	\boldsymbol{b}(\theta_l,r_l)) \otimes \boldsymbol{g}(\tau_l)  ) \in \mathbb{C}^{MPT}.
\end{align}
Thus, the log-likelihood function of $\boldsymbol{\xi}$ can be expressed as 
\begin{align}
	L(\boldsymbol{\xi})= -MPT\ln (\pi \sigma^2) -\frac{1}{\sigma^2}
	\left\|\boldsymbol{y}_v -\boldsymbol{\mu}(\boldsymbol{\xi})\right\|_F^2.
\end{align}

\subsection{Derivation of CRBs for Channel Parameters}
The Fisher Information matrix (FIM) for estimating the vector 
$\boldsymbol{\xi}$ is given by 
\begin{align}
	\boldsymbol{F}_{\boldsymbol{\xi}} &\triangleq \mathbb{E}\bigg\{
	\bigg(\frac{\partial L(\boldsymbol{\xi})}{\partial \boldsymbol{\xi}}\bigg)^H 
	\bigg(\frac{\partial L(\boldsymbol{\xi})}{\partial \boldsymbol{\xi}}\bigg)\bigg\}
	\nonumber\\
	&=\frac{2}{\sigma^2}
	\text{Re}\bigg\{
	\bigg(\frac{\partial \boldsymbol{\mu}(\boldsymbol{\xi})}{\partial \boldsymbol{\xi}}\bigg)^H
	\bigg(\frac{\partial \boldsymbol{\mu}(\boldsymbol{\xi})}{\partial \boldsymbol{\xi}}\bigg)\bigg\},
	\label{fim}
\end{align}
where $\frac{\partial \boldsymbol{\mu}(\boldsymbol{\xi})}{\partial \boldsymbol{\xi}}$ denotes 
the Jacobian matrix collecting the first-order partial derivatives of 
$\boldsymbol{\mu}(\boldsymbol{\xi})$ w.r.t. each parameter, 
i.e.,
\begin{align}
	\frac{\partial \boldsymbol{\mu}(\boldsymbol{\xi})}{\partial \boldsymbol{\xi}}\triangleq 
	\qquad \qquad \qquad \qquad \qquad \qquad \qquad \qquad \qquad \qquad&
	\nonumber\\
	\left[\frac{\partial \boldsymbol{\mu}(\boldsymbol{\xi})}{\partial \boldsymbol{\tau}} \
	\frac{\partial \boldsymbol{\mu}(\boldsymbol{\xi})}{\partial \boldsymbol{\theta}} \ 
	\frac{\partial \boldsymbol{\mu}(\boldsymbol{\xi})}{\partial \boldsymbol{r}} \ 
	\frac{\partial \boldsymbol{\mu}(\boldsymbol{\xi})}{\partial \boldsymbol{\alpha}_{R}} \ 
	\frac{\partial \boldsymbol{\mu}(\boldsymbol{\xi})}{\partial \boldsymbol{\alpha}_{I}} \right]
	\in \mathbb{C}^{MPT \times 5K}&,
	\nonumber\\
	\frac{\partial \boldsymbol{\mu}(\boldsymbol{\xi})}{\partial \boldsymbol{a}} \triangleq 
	\left[\frac{\partial \boldsymbol{\mu}(\boldsymbol{\xi})}{\partial a_1} \ \cdots \
	\frac{\partial \boldsymbol{\mu}(\boldsymbol{\xi})}{\partial a_K}\right] \in \mathbb{C}^{MPT\times K}&, 
	\nonumber\\
	\boldsymbol{a}\in\left\{\boldsymbol{\tau},\boldsymbol{\theta},\boldsymbol{r},
	\boldsymbol{\alpha}_{R},\boldsymbol{\alpha}_{I}\right\}&.
\end{align}
Specifically, we have 
\begin{align}
	&\frac{\partial \boldsymbol{\mu}(\boldsymbol{\xi})}{\partial \tau_l} = 
	\alpha_l \tilde{\boldsymbol{s}}_l \otimes (\boldsymbol{W}^H
	\boldsymbol{b}(\theta_l,r_l)) \otimes 
	\frac{\partial \boldsymbol{g}(\tau_l)}{\partial \tau_l}
	\nonumber\\
	&\frac{\partial \boldsymbol{\mu}(\boldsymbol{\xi})}{\partial \theta_l} = 
	\alpha_l \tilde{\boldsymbol{s}}_l \otimes \left(\boldsymbol{W}^H
	\frac{\partial \boldsymbol{b}(\theta_l,r_l)}{\partial \theta_l}\right) \otimes 
	\boldsymbol{g}(\tau_l),
	\nonumber\\
	&\frac{\partial \boldsymbol{\mu}(\boldsymbol{\xi})}{\partial r_l} = 
	\alpha_l \tilde{\boldsymbol{s}}_l \otimes \left(\boldsymbol{W}^H
	\frac{\partial \boldsymbol{b}(\theta_l,r_l)}{\partial r_l}\right) \otimes 
	\boldsymbol{g}(\tau_l),
	\nonumber\\
	&\frac{\partial \boldsymbol{\mu}(\boldsymbol{\xi})}{\partial \text{Re}\{\alpha_l\}} = 
	\tilde{\boldsymbol{s}}_l \otimes \left(\boldsymbol{W}^H\boldsymbol{b}(\theta_l,r_l)
	\right) 
	\otimes \boldsymbol{g}(\tau_l),
	\nonumber\\
	&\frac{\partial \boldsymbol{\mu}(\boldsymbol{\xi})}{\partial \text{Re}\{\alpha_l\}} = j
	\tilde{\boldsymbol{s}}_l \otimes (\boldsymbol{W}^H\boldsymbol{b}(\theta_l,r_l)) 
	\otimes \boldsymbol{g}(\tau_l),
\end{align}
where the derivative of the steering vectors are given by 
\begin{align}
	&\frac{\partial \boldsymbol{g}(\tau_l)}{\partial \tau_l} = 
	-j2\pi \text{diag}(f_1,\cdots,f_{P}) \boldsymbol{g}(\tau_l),
	\nonumber\\
	&\left[\frac{\partial \boldsymbol{b}(\theta_l,r_l)}{\partial \theta_l}\right]_{n}  = 
	j\frac{2\pi}{\lambda}\left(\frac{r_l(n-1)d\cos\theta_l}{r^{(n)}_l}\right)b_n,\ 
	\forall n,
	\nonumber\\
	&\left[\frac{\partial \boldsymbol{b}(\theta_l,r_l)}{\partial r_l}\right]_{n}  = 
	-j\frac{2\pi}{\lambda}\left(\frac{r_l - (n-1)d\sin\theta_l}{r^{(n)}_l}-1\right)b_n,\ 
	\forall n,
\end{align}
in which $\left[\frac{\partial \boldsymbol{b}(\theta_l,r_l)}{\partial \theta_l}\right]_{n}$, 
$\left[\frac{\partial \boldsymbol{b}(\theta_l,r_l)}{\partial r_l}\right]_{n}$ and
$b_n$ represent the $n$-th element of $\frac{\partial \boldsymbol{b}(\theta_l,r_l)}
{\partial \theta_l}$, $\frac{\partial \boldsymbol{b}(\theta_l,r_l)}{\partial r_l}$ 
and $\boldsymbol{b}(\theta_l,r_l)$, respectively. 

Then, the CRB for estimation error of $\boldsymbol{a}\in\left\{\boldsymbol{\tau},
\boldsymbol{\theta},\boldsymbol{r},\boldsymbol{\alpha}_{R},\boldsymbol{\alpha}_{I}
\right\}$ can be expressed as 
\begin{align}
	\text{MSE}^{\rm\small joint}(\boldsymbol{a}) &= \|\hat{\boldsymbol{a}} - \boldsymbol{a}\|_F^2 
	\nonumber\\ &\geq 
	\text{CRB}^{\rm\small joint}(\boldsymbol{a}) = \text{tr}\left([\boldsymbol{F}_{\boldsymbol{\xi}}^{-1}]
	_{\mathcal{I}_{a},\mathcal{I}_{a}}\right),
	\label{crb1}
\end{align}
where $\mathcal{I}_{a}$ is the index associated with the parameter $\boldsymbol{a}$. 
For example, $\mathcal{I}_{\theta} = [1,\cdots,K]$. 
\subsection{Derivation of CRBs for User Localization}
After obtaining the channel parameters $\{\hat{\theta}_k,\hat{\tau}_k\}_{k=1}^{K}$, the 
position of the $k$-th user can be calculated as 
\begin{align}
	\boldsymbol{p}_k\triangleq \left[{\hat{r}_k}\cos\hat{\theta}_k, 
	{\hat{r}_k}\sin\hat{\theta}_k \right]^T.
\end{align}  
Let $\boldsymbol{p} \triangleq \left[\boldsymbol{p}_1 \ \cdots \ 
\boldsymbol{p}_K\right]^T\in \mathbb{R}^{2K}$. 
The FIM for user localization can be obtained based on the FIM in (\ref{fim}), 
which is given by 
\begin{align}
	\boldsymbol{F}^{\rm\small joint}_{\boldsymbol{p}} = \boldsymbol{J}_{\boldsymbol{p}/\boldsymbol{\xi}}
	\boldsymbol{F}_{\boldsymbol{\xi}}
	\boldsymbol{J}_{\boldsymbol{p}/\boldsymbol{\xi}}^T,
\end{align}
where $\boldsymbol{J}_{\boldsymbol{p}/\boldsymbol{\xi}}$ denotes the partial derivative 
matrix w.r.t. $\boldsymbol{p}$, whose size is $2K\times 5K$, i.e.,
\begin{align}
	\boldsymbol{J}_{\boldsymbol{p}/\boldsymbol{\xi}} \triangleq 
	\left[\frac{\partial \boldsymbol{\theta}^T}{\partial \boldsymbol{p}} \ 
	\frac{\partial \boldsymbol{\tau}^T}{\partial \boldsymbol{p}} \ 
	\frac{\partial \boldsymbol{r}^T}{\partial \boldsymbol{p}}
	\frac{\partial \boldsymbol{\alpha}_R^T}{\partial \boldsymbol{p}}
	\frac{\partial \boldsymbol{\alpha}_I^T}{\partial \boldsymbol{p}}\right]^T \in \mathbb{C}^{5K \times 2K},
\end{align}
where 
\begin{align}
	&\frac{\partial \boldsymbol{a}^T}{\partial \boldsymbol{p}} \triangleq 
	\left[\frac{\partial a_1}{\partial \boldsymbol{p}} \ \cdots \ 
	\frac{\partial a_K}{\partial \boldsymbol{p}}\right]^T \in \mathbb{C}^{K \times 2K}, 
	\boldsymbol{a} \in \left\{\boldsymbol{\tau},\boldsymbol{\theta},\boldsymbol{r},
	\boldsymbol{\alpha}_{R},\boldsymbol{\alpha}_{I}\right\},
	\nonumber\\
	&\frac{\partial a_k}{\partial \boldsymbol{p}} = \left[\frac{\partial a_k}{\partial \boldsymbol{p}_1} 
	\ \cdots \ \frac{\partial a_k}{\partial \boldsymbol{p}_K}\right] \in \mathbb{C}^{2K},
	\forall k = 1,\cdots,K.
\end{align}
Specifically, we have 
\begin{align}
	&\frac{\partial \tau_k}{\partial \boldsymbol{p}_{k'}} = {\boldsymbol{0}_{2}, \forall k,k' = 1,\cdots,K,}
	\nonumber\\
	&\frac{\partial \theta_k}{\partial \boldsymbol{p}_{k'}} = \left\{\begin{array}{rcl}
		\boldsymbol{0}_{2}, &k \neq k',\\
		\left[\frac{\partial \theta_k}{p_{k,1}},\frac{\partial \theta_k}{p_{k,2}}\right]^T & k = k',
	\end{array}\right.
	\nonumber\\
	&\frac{\partial r_k}{\partial \boldsymbol{p}_{k'}} =  {\left\{\begin{array}{rcl}
		\boldsymbol{0}_{2}, &k \neq k',\\
		\left[\frac{\partial r_k}{p_{k,1}},\frac{\partial r_k}{p_{k,2}}\right]^T & k = k',
	\end{array}\right. }
	\nonumber\\
	&\frac{\partial \text{Re}\{\alpha_k\}}{\partial \boldsymbol{p}_{k'}} = \boldsymbol{0}_{2}, \forall k,k' = 1,\cdots,K,
	\nonumber\\
	&\frac{\partial \text{Im}\{\alpha_k\}}{\partial \boldsymbol{p}_{k'}} = \boldsymbol{0}_{2}, \forall k,k' = 1,\cdots,K.
\end{align}
To compute ${\left\{\frac{\partial r_k}{p_{k,i}},\frac{\partial \theta_k}{p_{k,i}}\right\}},
{i=1,2}$, we can first compute
\begin{align}
	{
	\frac{\partial (p_{k,1},p_{k,2})}{\partial (r_k,\theta_k)} = \begin{bmatrix}
		&\cos\theta_k &-r_k\sin\theta_k\\
		&\sin\theta_k & r_k\cos\theta_k \\
	\end{bmatrix}.}
\end{align}
Consequently, it can be derived that 
{
\begin{align}
	\frac{\partial (r_k,\theta_k)}{(p_{k,1},p_{k,2})} = 
	\left[\frac{\partial (p_{k,1},p_{k,2})}{\partial (r_k,\theta_k)}\right]^{-1}=
	\begin{bmatrix}
		&\cos\theta_k &\sin \theta_k \\
		&\frac{-\sin\theta_k}{r_k} &\frac{\cos \theta_k}{r_k} \\
	\end{bmatrix}.
\end{align}}
Finally, the CRBs for user localization is given by
\begin{align}
	\text{MSE}^{\rm\small joint}(\boldsymbol{p}) &= \|\boldsymbol{p}-\hat{\boldsymbol{p}}\|_F^2
	\nonumber\\&\geq \text{CRB}^{\rm\small joint}(\boldsymbol{p}) = 
	\text{tr}\left((\boldsymbol{F}_{\boldsymbol{p}}^{\rm\small joint})^{-1}\right).
\end{align}

{
\section{Derivation of CRBs for CPD-Delay-Aided Method}
For the CPD-Delay-Aided method, the propagation delay is first estimated from the recovered
factor matrix $\hat{\boldsymbol{G}}_o$, and the distance is subsequently recovered according to
the relationship $\hat{r}_k = c \tau_{k},\forall k = 1,\cdots,K$. Since the delay estimation 
relies on the same parameter extraction procedure as that of the CPD-Joint method, the CRB 
associated with $\boldsymbol{\tau}$ remains identical to the CRB derived in \eqref{crb1}. 
For the other parameters, the delay and distance should be represented by a single parameter 
and the parameter vector is reduced to 
$\tilde{\boldsymbol{\xi}} \triangleq \left[\boldsymbol{\theta}^T \ \boldsymbol{r}^T \ 
\boldsymbol{\alpha}_{R}^T \ \boldsymbol{\alpha}_{I}^T\right]^T$. 

In this case, the noiseless part of $\boldsymbol{y}_v$ can be re-written as 
\begin{align}
	\boldsymbol{\mu}(\tilde{\boldsymbol{\xi}}) \triangleq \sum\nolimits_{l=1}^{L}
	\alpha_l (\tilde{\boldsymbol{s}}_l \otimes (\boldsymbol{W}^H 
	\boldsymbol{b}(\theta_l,r_l)) \otimes \boldsymbol{g}(r_l/c) ) \in \mathbb{C}^{MPT}.
\end{align}

Accordingly, the FIM for estimating the reduced vector $\tilde{\boldsymbol{\xi}}$ is given by 
\begin{align}
	\boldsymbol{F}_{\tilde{\boldsymbol{\xi}}} 
	&=\frac{2}{\sigma^2}
	\text{Re}\bigg\{
	\bigg(\frac{\partial \boldsymbol{\mu}(\tilde{\boldsymbol{\xi}})}{\partial \tilde{\boldsymbol{\xi}}}\bigg)^H
	\bigg(\frac{\partial \boldsymbol{\mu}(\tilde{\boldsymbol{\xi}})}{\partial \tilde{\boldsymbol{\xi}}}\bigg)\bigg\},
	\label{fim2}
\end{align}
where 
\begin{align}
	\frac{\partial \boldsymbol{\mu}(\tilde{\boldsymbol{\xi}})}{\partial \tilde{\boldsymbol{\xi}}}\triangleq 
	\left[
	\frac{\partial \boldsymbol{\mu}(\boldsymbol{\xi})}{\partial \boldsymbol{\theta}} \ 
	\frac{\partial \boldsymbol{\mu}(\boldsymbol{\xi})}{\partial \boldsymbol{r}} \ 
	\frac{\partial \boldsymbol{\mu}(\boldsymbol{\xi})}{\partial \boldsymbol{\alpha}_{R}} \ 
	\frac{\partial \boldsymbol{\mu}(\boldsymbol{\xi})}{\partial \boldsymbol{\alpha}_{I}} \right]
	\in \mathbb{C}^{MPT \times 5K}.
\end{align}

Then, we compute the partial derivatives of $\boldsymbol{\mu}(\tilde{\boldsymbol{\xi}})$ 
w.r.t. each parameter. For the distance parameter $\boldsymbol{r}$, we have 
\begin{align}
	\frac{\partial \boldsymbol{\mu}(\tilde{\boldsymbol{\xi})}}{\partial r_l} 
	&=\alpha_l \tilde{\boldsymbol{s}}_l \otimes \left(\boldsymbol{W}^H
	\frac{\partial \boldsymbol{b}(\theta_l,r_l)}{\partial r_l}\right) \otimes 
	\boldsymbol{g}(r_l/c)
	\nonumber\\
	&\quad +\alpha_l\tilde{\boldsymbol{s}}_l \otimes (\boldsymbol{W}^H
	\boldsymbol{b}(\theta_l,r_l)) \otimes 
	\frac{\partial \boldsymbol{g}(r_l/c)}{\partial r_l},
\end{align}
which consists of two parts: the direct dependence on $\boldsymbol{r}$ though the 
near-field steering vector, and the indirect dependence on $\boldsymbol{r}$ 
through the delay $\tau_l = r_l/c$. 
Specifically, $\frac{\partial \boldsymbol{g}(r_l/c)}{\partial r_l}$ is given by 
\begin{align}
	\frac{\partial \boldsymbol{g}(r_l/c)}{\partial r_l} =
	-j\frac{2\pi}{c} \text{diag}(f_1,\cdots,f_{P}) \boldsymbol{g}(r_l/c),
\end{align}
Moreover, the partial derivatives of $\boldsymbol{\mu}(\tilde{\boldsymbol{\xi}})$ w.r.t. 
the other parameters are the same as that of $\boldsymbol{\mu}(\boldsymbol{\xi})$, 
i.e., 
\begin{align}
	\frac{\partial \boldsymbol{\mu}(\tilde{\boldsymbol{\xi})}}{\partial a_l}
	=\frac{\partial \boldsymbol{\mu}(\boldsymbol{\xi})}{\partial a_l}, 
	\quad a =\{\theta,\alpha\}, \quad \forall l,
\end{align}

Consequently, for any parameter block $\boldsymbol{a}\in\{
\boldsymbol{\theta},\boldsymbol{r},\boldsymbol{\alpha}_{R},\boldsymbol{\alpha}_{I}\}$, the corresponding CRB is given by 
\begin{align}
	\text{MSE}^{\rm\small Delay}(\boldsymbol{a}) &= \|\hat{\boldsymbol{a}} - \boldsymbol{a}\|_F^2 
	\nonumber\\&\geq 
	\text{CRB}^{\rm\small Delay}(\boldsymbol{a}) = \text{tr}\left([\boldsymbol{F}_{\tilde{\boldsymbol{\xi}}}^{-1}]
	_{\mathcal{I}_{a},\mathcal{I}_{a}}\right).
\end{align}

Following the similar steps in Appendix A-B, the CRBs for user localization can be obtained 
by 
\begin{align}
	\text{MSE}^{\rm\small Delay}(\boldsymbol{p}) &= \|\boldsymbol{p}-\hat{\boldsymbol{p}}\|_F^2
	\nonumber\\ &\geq \text{CRB}^{\rm\small Delay}(\boldsymbol{p}) = 
	\text{tr}\left((\boldsymbol{F}_{\boldsymbol{p}}^{\rm delay})^{-1}\right),
\end{align}
where 
\begin{align}
	&\boldsymbol{F}_{\boldsymbol{p}}^{\rm\small Delay} = \boldsymbol{J}_{\boldsymbol{p}/\tilde{\boldsymbol{\xi}}}
	\boldsymbol{F}_{\tilde{\boldsymbol{\xi}}}
	\boldsymbol{J}_{\boldsymbol{p}/\tilde{\boldsymbol{\xi}}}^T,
	\nonumber\\
	&\boldsymbol{J}_{\boldsymbol{p}/\tilde{\boldsymbol{\xi}}} = 
	\boldsymbol{J}_{\boldsymbol{p}/\boldsymbol{\xi}}(K+1:5K,:).
\end{align}
}
\end{appendices}

\bibliography{newbib}
\bibliographystyle{IEEEtran}

\end{document}